\documentclass{article}
\usepackage[left=3cm,right=3cm,top=3cm,bottom=3cm]{geometry} 
\usepackage{amsmath} 
\usepackage{authblk}

\RequirePackage[OT1]{fontenc}
\RequirePackage{amsthm,amsmath}
\RequirePackage[numbers]{natbib}
\RequirePackage[colorlinks,citecolor=blue,urlcolor=blue]{hyperref}
\usepackage{amsmath} 

\usepackage{marvosym}
\usepackage{amssymb}
 \usepackage{amsmath}
 \usepackage{natbib}
 \usepackage{float}
 \usepackage{rotating}
 \usepackage{amsmath}
 \usepackage{subfig}
 \usepackage{multirow}
 \usepackage{float}
 \usepackage{mathtools}
 \mathtoolsset{showonlyrefs=true}
\newcommand{\bSigma}{ \mbox{\boldmath $\Sigma$} }
\newcommand{\bA}{ \mbox{$\mathbf{A}$ }}
\newcommand{\ba}{ \mbox{$\mathbf{a}$ }}
\newcommand{\bw}{ \mbox{$\mathbf{w}$ }}

\newcommand{\bs}{ \mbox{\boldmath $s$}}

\usepackage{xcolor}

\begin{document}

\title{A Hierarchical Multivariate Spatio-Temporal Model for Large Clustered Climate data with Annual Cycles
}


\author[1]{Gianluca Mastrantonio}
\author[2]{Giovanna Jona Lasinio}
\author[3]{Alessio Pollice}
\author[4]{Giulia Capotorti}
\author[5]{Lorenzo Teodonio}
\author[4]{Giulio Genova}
\author[4]{Carlo Blasi}

\affil[1]{Department of Mathematical Science, \textbf{ Politecnico di Torino}}
\affil[2]{Department of Statistical Sciences,  \textbf{Sapienza Universit\`a di Roma}}
\affil[3]{Department of Economics and Finance, Aldo Moro \textbf{Universit\`a di  Bari}}
\affil[4]{Department of Environmental Biology, \textbf{Sapienza Universit\`a di Roma}}
\affil[5]{ICRCPAL, Ministry of Cultural Heritage and Activities and Tourism, Roma}
\date{   }

\maketitle

\begin{abstract}
We present a multivariate hierarchical space-time model to describe the joint series of monthly extreme temperatures and amounts of
rainfall. Data are available for 360 monitoring stations over 60 years, with missing data affecting almost all series. Model components
account for spatio-temporal dependence with annual cycles, dependence on covariates and between responses. The very large amount of data is
tackled modeling the spatio-temporal dependence by the nearest neighbor Gaussian process. Response multivariate dependencies are described using
the linear model of coregionalization, while annual cycles are assessed by a circular representation of time. The proposed approach allows
imputation of missing values and easy interpolation of climate surfaces at the national level. The motivation behind is the
characterization of the so called ecoregions over the Italian territory.  Ecoregions delineate broad and discrete ecologically homogeneous
areas of similar potential as regards the climate, physiography, hydrography, vegetation and wildlife, and provide a geographic framework
for interpreting ecological processes, disturbance regimes, vegetation patterns and dynamics. To now, the two main Italian macro-ecoregions
are hierarchically arranged into 35 zones. The current climatic characterization of Italian ecoregions is based on data and bioclimatic
indices for the period 1955-1985 and requires an appropriate update.
\end{abstract}

\section{Introduction}\label{cyc:intro}
Climate elements and regimes, such as temperature, precipitation and their annual cycles, primarily affect type and distribution of plants,
animals, and soils as well as their combination in complex ecosystems \citep{Bailey2004, Metzger2013}. As such, the ecological
classification of climate represents one of the basic step for the definition and mapping of ecoregions, i.e. of broad ecosystems occurring in
discrete geographical areas \citep{bailey1983, loveland2004}. In keeping with these assumptions, a hierarchical classification of the
ecoregions was recently performed in Italy including climate among the main diagnostic features, together with biogeography and physiography \citep{blasietal2014}. The Italian
ecoregions (see figure \ref{fig:ecoregionslev}) are arranged into four hierarchically nested tiers, which consist of 2 Divisions, 7 Provinces, 11 Sections, and 33 Subsections\footnote{\textbf{1 Temperate Division}. 1A Alpine Province. 1A1 Western Alps Section; 1A1a Alpi Marittime Subsection; 1A1b Northwestern Alps Subsection; 1A2 Central and Eastern Alps Section; 1A2a Pre-Alps Subsection; 1A2b Dolomiti and Carnia Subsection; 1A2c Northeastern Alps Subsection. 1B Po Plain Province; 1B1 Po Plain Section; 1B1a Lagoon Subsection; 1B1b Central Plain Subsection; 1B1c Western Po Basin Subsection. 1C Apennine Province; 1C1 Northern and Western Apennine Section; 1C1a Toscana and Emilia-Romagna Subsection; 1C1b Tuscan Basin Subsection; 1C2 Central and Southern Apennine Section; 1C2a Umbria and Marche Apennine Subsection; 1C2b Lazio and Abruzzo Apennine Subsection; 1C2c Campania Apennine Subsection; 1D Italian part of Illyrian Province.\\
\textbf{2 Mediterranean Division}. 2A Italian part of Ligurian-Provencal Province. 2B Tyrrhenian Province; 2B1 Northern and Central Tyrrhenian Section;2B1a Eastern Liguria Subsection; 2B1b Maremma Subsection; 2B1c Roman Area Subsection; 2B1d Southern Lazio Subsection; 2B2 Southern Tyrrhenian Section; 2B2a Western Campania Subsection; 2B2b Lucania Subsection; 2B2c Cilento Subsection; 2B2d Calabria Subsection; 2B3 Sicilia Section; 2B3a Iblei Subsection; 2B3b Sicilia Mountains Subsection; 2B3c Central Sicilia Subsection; 2B3d Western Sicilia Subsection; 2B4 Sardegna Section; 2B4a Southwestern Sardegna Subsection; 2B4b Northwestern Sardegna Subsection; 2B4c Southeastern Sardegna Subsection; 2B4d Northeastern Sardegna Subsection. 2C Adriatic Province; 2C1 Central Adriatic Section; 2C1a Abruzzo and Molise Adriatic Subsection; 2C1b Marche Adriatic Subsection; 2C2 Southern Adriatic Section; 2C2a Murge and Salento Subsection; 2C2b Gargano Subsection.}.
For each of the tiers a different climatic diagnostic detail has been adopted: different macroclimatic zones and regions characterize
Divisions and Provinces, whereas different bioclimatic types and ranges in observed
thermo-pluviometric data characterize Section and Subsections.\\

\subsection{Bioclimatic classification}
To now, the climatic features adopted for the diagnosis and description of the Italian ecoregions mainly refer to data and bioclimatic
indices that date back to the period 1955-1985. It needs to be updated in order to validate ecoregion boundaries, summarize current and past climatic conditions of the
ecoregions, assess climate impacts on ecosystems at the meso-scale and formulate reliable biodiversity conservation strategies. To these
aim several models and approaches have been proposed \citep[see][]{Hijmans2005, Pesaresi2014} and several limitations have
been pointed out such over smoothing in the interpolation of climate variables, severe loss of precision at small spatial resolution and
more. As a matter of fact, temperature and precipitation data are generally heavily discretized in both space and time and require the
interpolation to climate surfaces. Interpolations are usually carried out neglecting the correlation between climatic variables and provide
climate surfaces that are often over-smoothed especially at the sub-continental scale. The uncertainty of estimated temperatures and
precipitations considerably increases in areas characterized by large variation in elevation or with sparse weather monitoring stations.
\cite{Gopar2015} clearly state that ``rigorous mapping of climatic patterns outstands as one of the mayor issues concerning climatic
change''. In their paper they investigate the extent of the bioclimatic approach to develop a rigorous cartographic methodology to express
climatic diversity patterns. Their work is strongly affected by the quality of climate surfaces available at the chosen scale.
\cite{McKenney2010} present four approaches that summarize projected climate changes across Ontario's ecosystems at two spatial scales
(ecoregions and selected natural heritage areas). The four approaches are based on interpolated climate
surfaces obtained neglecting the correlation among climatic variables and without a rigorous assessment of estimates uncertainty. 
The majority of the above cited works define and analyze ecoregions making extensive use of the WorldClim database.
%
%
The most recent release of the WorldClim database was obtained using the work of \cite{fick2017} who present an updated version of an older
protocol due to \cite{Hijmans2005}. With this new release WorldClim includes independent spatially interpolated monthly estimates of many
climate variables for global land areas, at approximately 1 $km^2$ spatial resolution. Monthly values of temperature (minimum, maximum and average),
precipitation, solar radiation, vapor pressure and wind speed are aggregated across the target temporal range 1970-2000 using data from
between 9000 and 60000 weather monitoring stations. Weather station data were interpolated using thin-plate splines with covariates
including elevation, distance to the coast and three satellite-derived variables: maximum and minimum land surface temperature as well as
cloud cover obtained by the MODIS satellite platform. The authors propose to use a multi-step procedure, adopting the best performing model
for each region and variable. Although this solution allows for an improvement in terms of goodness of fit, it does not allow for an easy
evaluation of the overall uncertainty and does not avoid the risk of over-smoothing that was already observed with the previous protocol.

\subsection{The available data}\label{cyc:data}
Precipitation and min/max temperature data were recorded monthly at 360 monitoring stations over 60
years (1951-2010). Therefore, the overall database consists of approximately 750000 records for three variables. The data were mostly
obtained from National Institutions, such as ISPRA (``Progetto Annali'' and SCIA), CRA/CREA, Meteomont (Guardia Forestale) and ENEA.
Furthermore, data from additional stations were acquired from numerous Italian local authorities (Regions, Provinces)\footnote{Data
sources, organized by region: Abruzzo: Regione Abruzzo, direzione Lavori Pubblici e Protezione Civile;
 Basilicata: Regione Basilicata, Ufficio Protezione Civile;
Calabria:  Regione Calabria, ARPACAL, Centro funzionale multi-rischi; Campania: Regione Campania, Direzione generale Protezione Civile;
Emilia Romagna: ARPA Emilia Romagna:
 Friuli Venezia Giulia: ARPA Friuli Venezia Giulia, Protezione Civile Regionale;
 Lazio: Regione Lazio, Servizio Integrato Agrometeorologico;
 Liguria: Arpa Liguria;
 Lombardia: Arpa Lombardia, Protezione Civile Regionale;
 Marche: Regione Marche, Servizio Agrometeo Regionale;
 Molise: Protezione Civile Regionale;
 Piemonte: Arpa Piemonte;
 Puglia: Regione Puglia, Arpa, Protezione Civile Regionale;
 Sardegna: Regione Sardegna, Arpa Sardegna;
 Sicilia: Regione Sicilia, Osservatorio Acque, Assessorato dell'Energia e dei servizi di pubblica utilit\`a, dipartimento dell'Acqua e dei rifiuti;
 Trentino: Provincia di Trento, Centro funzionale Protezione Civile;
 Toscana: Regione Toscana, Settore idrologico regionale;
 Umbria: Regione Umbria, Centro funzionale decentrato di monitoraggio meteo-idrologico;
 Valle d'Aosta: Arpa Valle D'Aosta;
 Veneto: Arpa Veneto.
}. 
Almost all time series are affected by variable amounts of missing data as shown in table \ref{tab:missingdata} reporting summary statistics on percentages of missing values by stations.
\begin{table}
\begin{center}
\begin{tabular}{|l|cccccc|}
\hline &Min. &1st Qu.&  Median &   Mean& 3rd Qu.&    Max.\\ \hline
 Rain& 0.00 &  1.88&   7.64&  11.50&  18.47&  68.33\\
 T. max& 0.00&   5.42&  13.47&  16.50&  24.31&  96.25\\
 T. min&  0.00 &  5.45&  13.47&  16.52&  24.17 & 96.25\\\hline
 \end{tabular}
 \caption{Summary statistics of the percentages of missing data at monitoring station.}\label{tab:missingdata}
  \end{center}
 \end{table}
The observed climate variables vary consistently with the 33 ecoregional subsections, as is shown in Figure \ref{fig:boxtutti}.




\begin{figure}[t]
\centering
\subfloat[$k=5$ ]{\includegraphics[width=9cm,height=7cm]{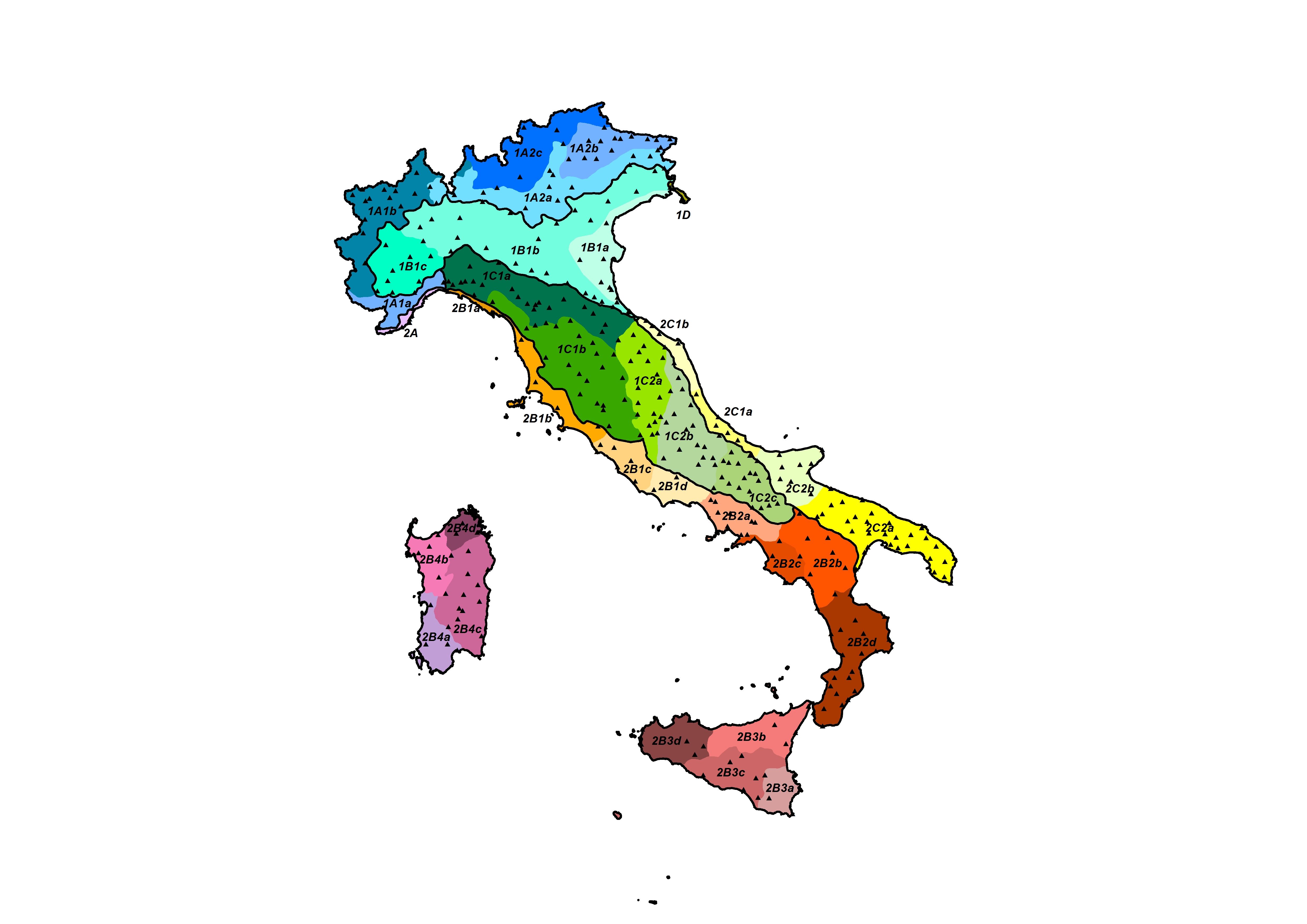}}\\
\subfloat[$k=2$]{\includegraphics[width=4cm,height=3cm]{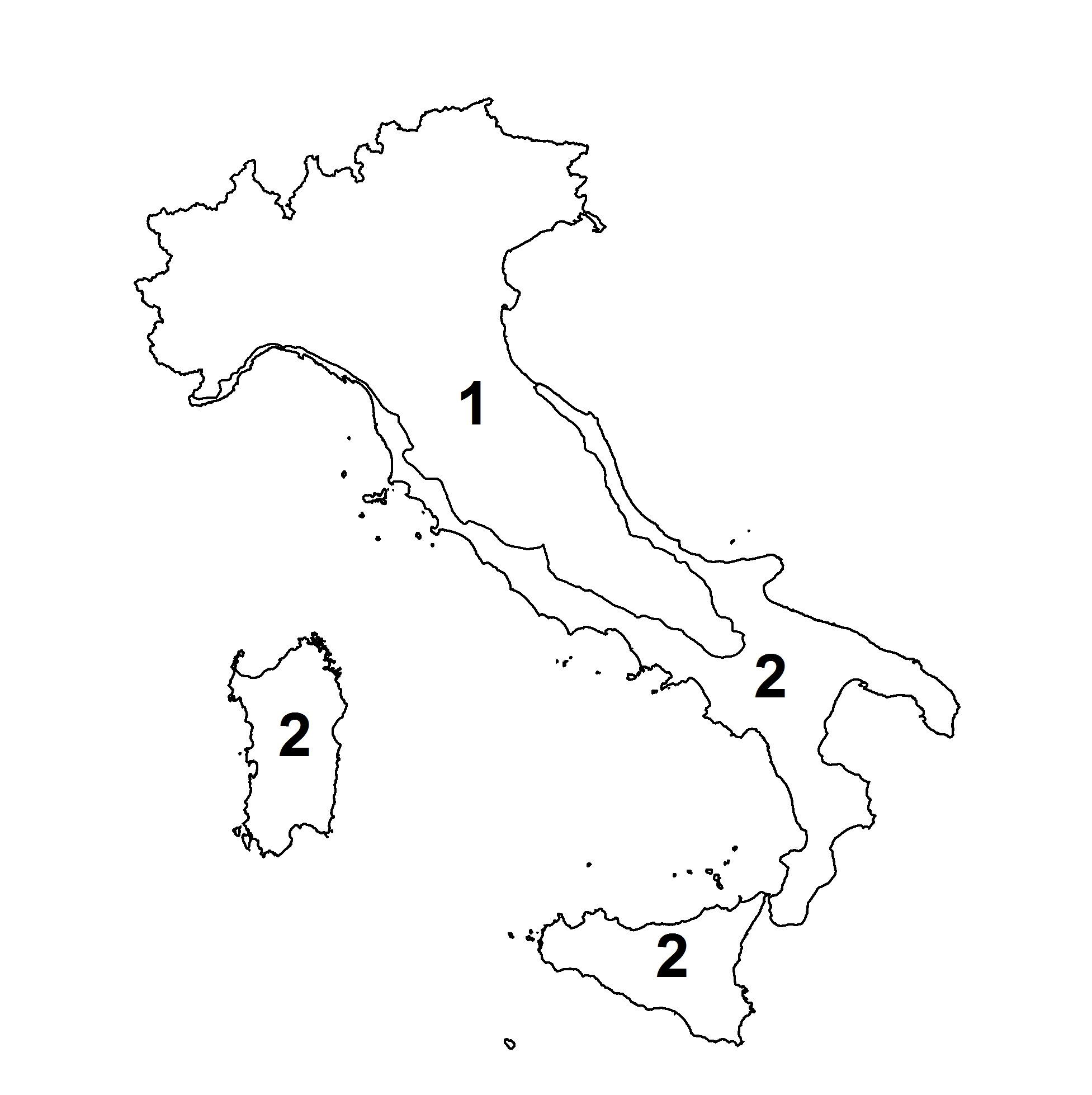}} \subfloat[$k=3$]{\includegraphics[width=4cm,height=3cm]{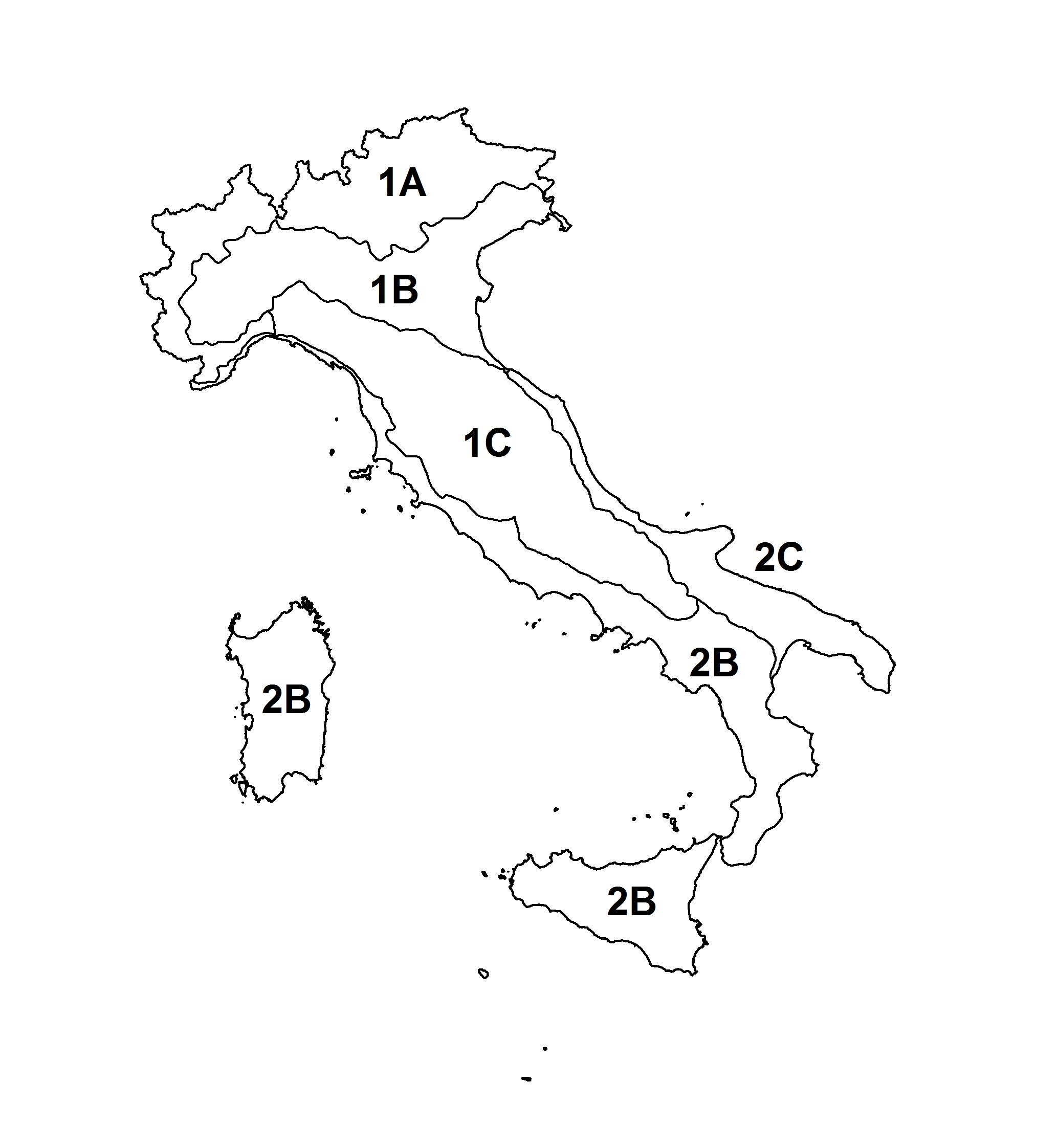}} \subfloat[$k=4$]{\includegraphics[width=4cm,height=3cm]{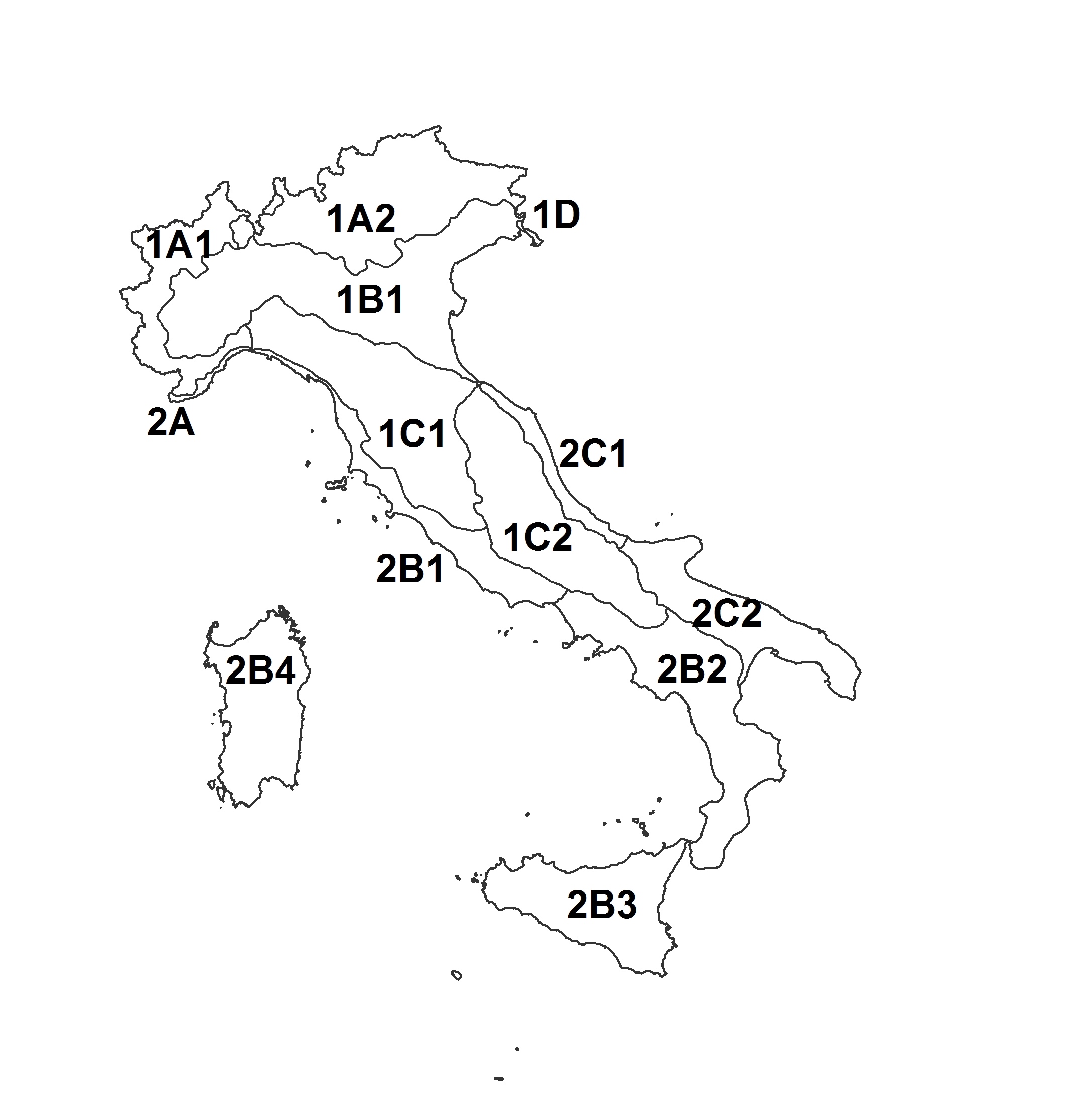}}\caption{Ecoregional hierarchical tier organization and climate monitoring network (a).}\label{fig:ecoregionslev}
\end{figure}


\begin{figure}
\centering
\subfloat[Maximum Temperature]{\includegraphics[width=4.5cm,height=4cm]{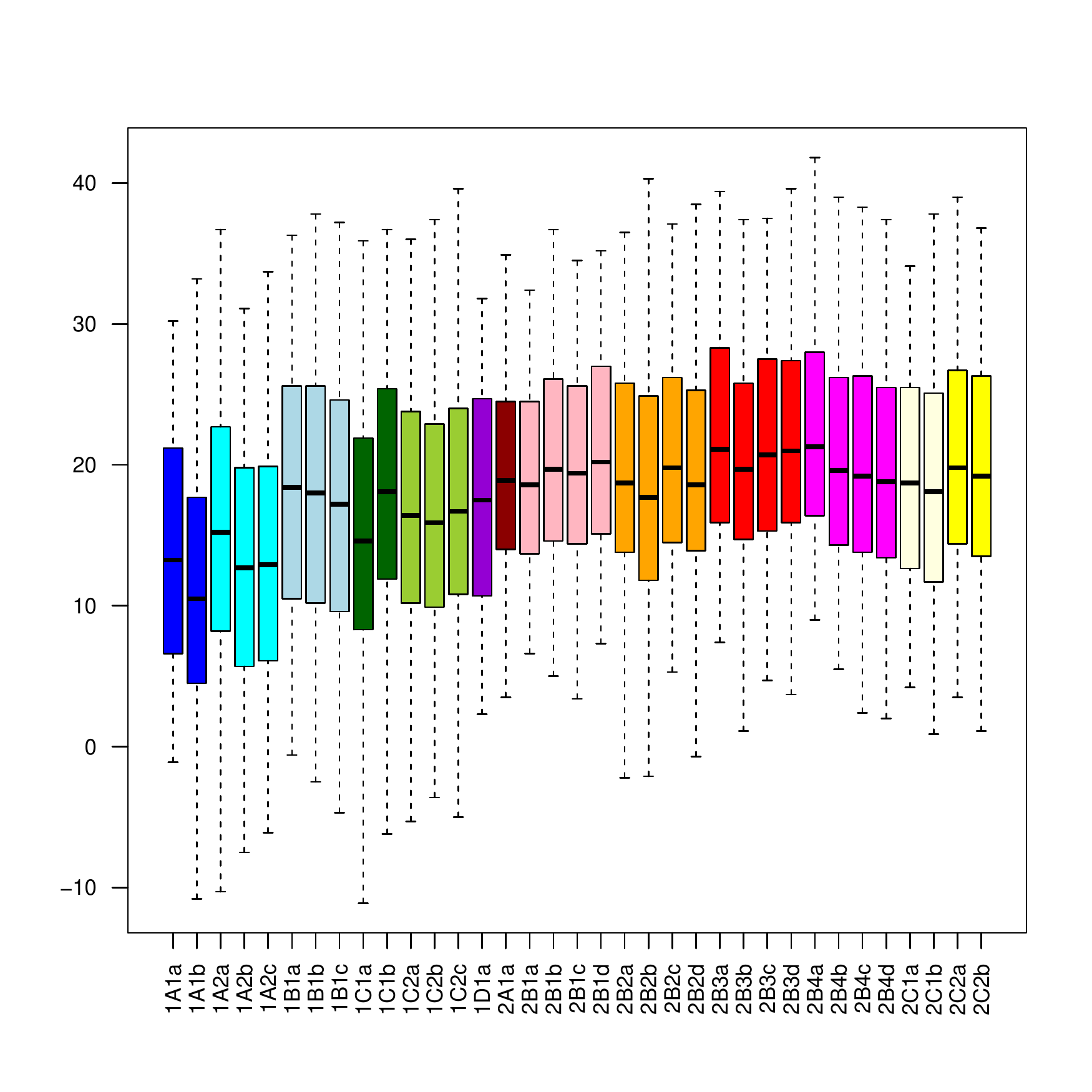}}\subfloat[Minimum Temperature]{\includegraphics[width=4.5cm,height=4cm]{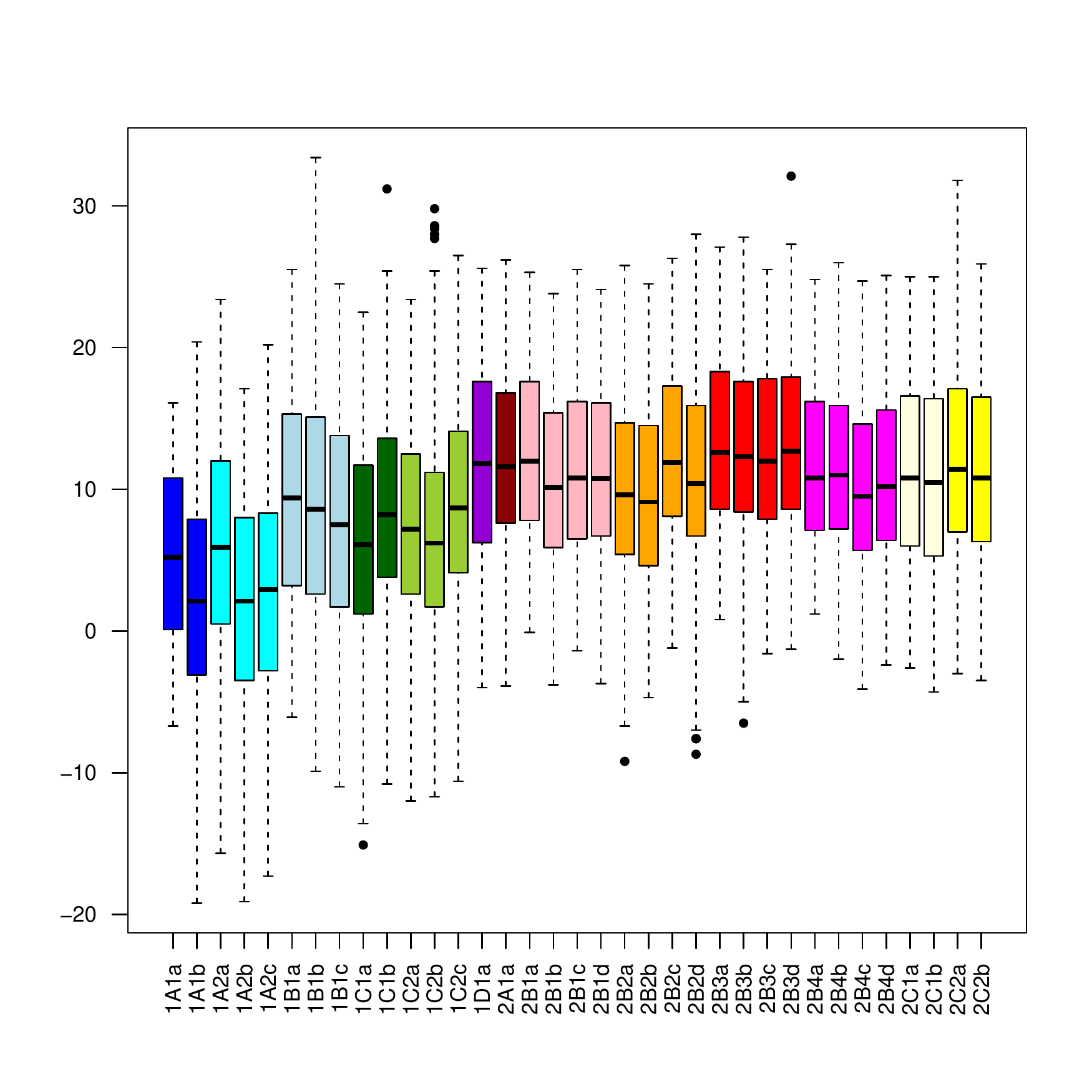}}\subfloat[Log-precipitation]{\includegraphics[width=4.5cm,height=4cm]{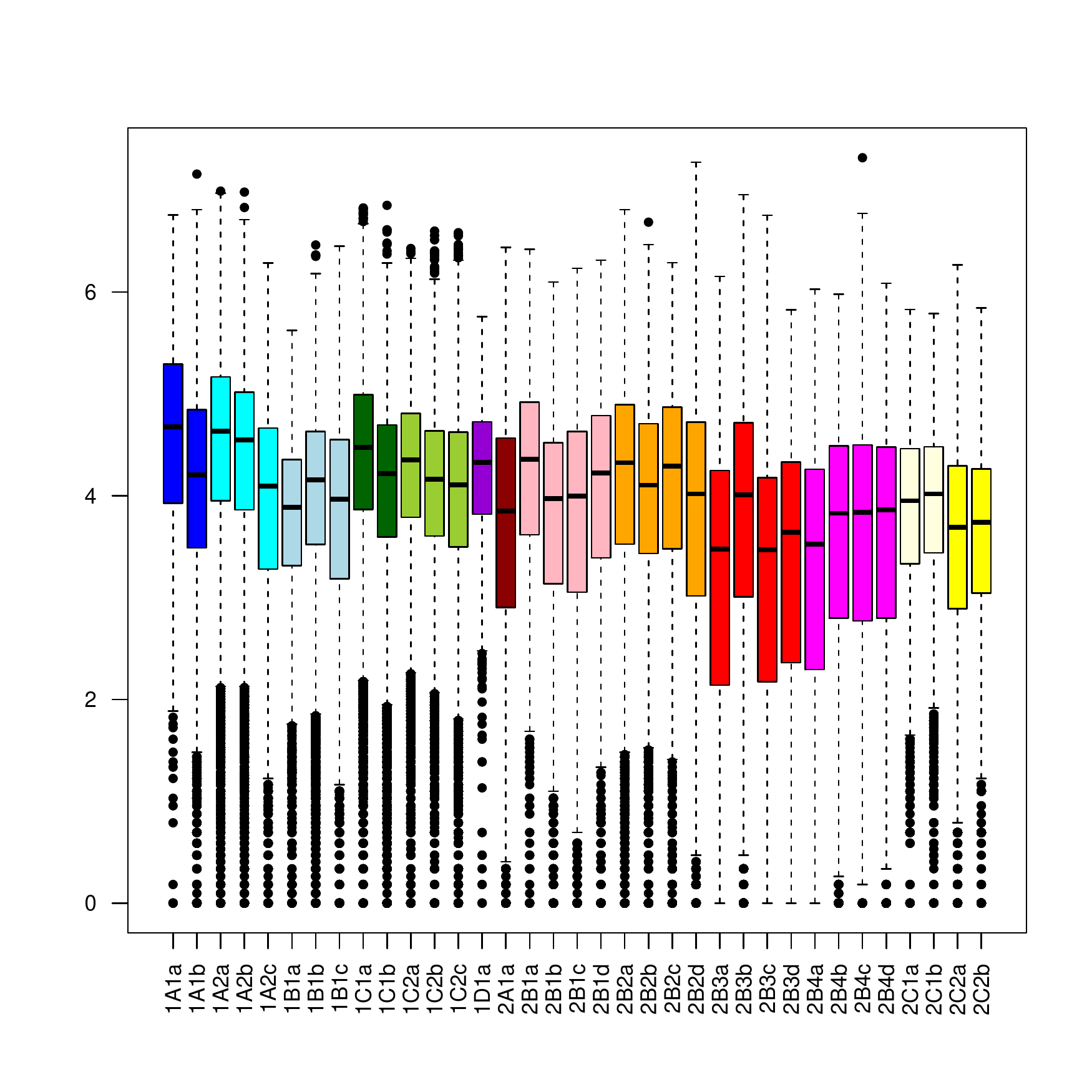}}

\caption{Boxplots of maximum, minimum temperature and rainfall by ecoregions. Colors follow the Italian ecoregional Sections, see figure \ref{fig:ecoregionslev}.}\label{fig:boxtutti}
\end{figure}

\subsection{Spatio-temporal interpolation of large datasets}
Bioclimatic classification requires an effective interpolation approach accounting for the correlation among climate variables
and such that uncertainty evaluation is rigorously obtained.   A very large amount of literature on the interpolationj of massive spatial and spatio-temporal data is now available, with some review papers \citep{Jona2013b, LI2016} and
books \citep{Banerjee2014,gelfand2010, cressie2011} to which the interested reader is referred for details. In what follows, we restrict
our interest to inferences that can be carried on at arbitrary spatial and temporal resolutions,
possibly finer than those of the observed data, 
and to approaches that allow for a rigorous evaluation of the overall uncertainty
and that are computationally feasible. The common choice would be to construct a stochastic process model to capture dependence using a
spatio-temporal covariance function \citep[][and references therein]{gelfand2010}. While the richness and flexibility of spatio-temporal
process models are indisputable, their computational feasibility and implementation pose major challenges for large datasets. Model-based
inference almost always involves the calculation of inverses  of large matrices, say $N\times N$, where $N$ is the number
of spatio-temporal points considered. Unless the matrix is sparse or has a specific structure that eases the computation,
$\mathcal{O}(N^3)$ floating points operations would be required for any matrix inversion. Approaches for modeling large covariance matrices
in purely spatial settings include low rank and covariance tapering models \citep[][and references therein]{gelfand2010, Banerjee2014,
cressie2011} and multivariate tapering proposed in \citet{Bevi2016a}, approximations using Gaussian Markov Random Fields (GMRF), the
Laplace transform and Stochastic Partial differential Equations \citep{rueheld2005, Rue2009, Lindgren2011,BLANGIARDO2013,
blangiardocameletti2015}, products of lower dimensional conditional densities \citep[see][and references therein]{datta2016, datta2016a}
and composite likelihoods \citep[][]{Eidsvik2014} with the recent multivariate extension proposed in \citet{Bevi2016}. A large number of
extensions to spatio-temporal settings have been proposed, including \cite{Cressie2010}, \cite{Finley2012} and \cite{Katzfuss2012} who
introduce dynamic spatio-temporal low-rank spatial processes, while \cite{Xu2015} choose  a GMRF approach. All the previous works use
dynamic models defined for fixed temporal lags and are not easily extended to continuous spatio-temporal domains. Continuous
spatio-temporal process modeling of large data has received relatively less attention. Composite likelihoods are proposed in \cite{Bai2012}
and \cite{Bevilacqua2012} for parameter estimation in a continuous space-time setup. Both papers focus upon constructing computationally
feasible likelihood approximations and inference is restricted  to parameter estimation only. Uncertainty estimates are mostly based on
asymptotic results which are often inappropriate for irregularly observed data. Moreover, in both cases predictions at arbitrary locations
and time points are obtained imputing estimates into an interpolator derived with a different process model. Furthermore, computations are
expensive for large $N$ and may not be accurate in reflecting the predictive uncertainty. Finally, an interesting proposal was recently
provided by \cite{datta2016} with the definition of the Nearest Neighbor Gaussian Process (NNGP) approximation for large continuous spatial
datasets, extended to the spatio-temporal framework in \cite{datta2016a}.

In this work we are going to apply, introducing several novelties, the approach proposed by \cite{datta2016a} to a multivariate spatio-temporal setting,
using a generalised  NNGP approximation on a multi-response problem. To this aim, we combine the NNGP with the linear model of
coregionalization \citep{Gelfand2004} and with a circular representation of time \citep[see for instance][]{shirota2017} to include annual
cycles. We obtain a computationally feasible tool that allows for multivariate interpolation with large continuous space-time data,
providing an accurate evaluation of the associated uncertainties. The proposed approach is applied to the characterization of Italian
ecoregions in terms of temperature and precipitation, providing successful answers to the problems mentioned above: we estimate a joint
model that accounts for climate variables correlation and obtain rigorous assessment of all estimates and predictions uncertainty. We deal
with a huge amount of data and avoid over-smoothing. Further, we introduce annual cycles using a novel and flexible tool, we discuss the
``best'' ecoregions hierarchical classification tier in terms of temperature and precipitation characterization and, eventually, we get a
direct imputation of missing data.


While in section \ref{cyc:data} we gave a full description of the available data and of the four ecoregion hierarchical classification
tiers, the rest of the paper is organized as follows. Section \ref{cyc:model} is dedicated to the definition of the multivariate
coregionalization model for the Italian data, while section \ref{cyc:imp} contains the NNGP definition and some details of the
implementation of estimates and predictions. Results are reported and commented in section \ref{cyc:result}, while section
\ref{cyc:conclusion} contains some final remarks and addresses for future developments.

\section{The Model}\label{cyc:model}

Let $\bs \in \mathcal{S} \subset \mathbb{R}^d$, with $d=2$, and $t \in \mathcal{T} \subset \mathbb{R}$ be spatial and temporal coordinates
respectively, and let $Y^*_{1}(\bs,t)$, $Y^*_{2}(\bs,t)$ and $Y^*_{3}(\bs,t)$ represent the precipitation level, minimum and maximum
temperatures observed at $(\bs,t)$. Then these variables have the following constraints: $Y^*_{1}(\bs,t) \geq 0 $ and $Y^*_3(\bs,t)\geq
Y^*_2(\bs,t)$. To simplify modeling and computations, we prefer to work with latent variables defined over the entire real line
$\mathbb{R}$, embedding the above constraints in the variable definitions. Latent variables $Y_{1}(\bs,t)$, $Y_{2}(\bs,t)$ and
$Y_{3}(\bs,t)$ are defined as follows:
\begin{equation}
\begin{cases}
Y_{1}(\bs,t) = Y^*_{1}(\bs,t) & \mbox{ if }  Y^*_{1}(\bs,t)  >0,\\
Y_{1}(\bs,t) \le0 & \mbox{ if }  Y^*_{1}(\bs,t)  =0,
\end{cases} \label{eq:y1}
\end{equation}
$$Y_{2}(\bs,t) = Y^*_{2}(\bs,t),$$
\begin{equation}
\begin{cases}
Y_{3}(\bs,t) = Y^*_{3}(\bs,t) - Y^*_{2}(\bs,t)& \mbox{ if }  Y^*_{3}(\bs,t) - Y^*_{2}(\bs,t)  >0,\\
Y_{3}(\bs,t) \le 0 & \mbox{ if }  Y^*_{3}(\bs,t) - Y^*_{2}(\bs,t)=0.
\end{cases}\label{eq:y3}
\end{equation}
%
Each latent response $Y_i$, $i=1,2,3$ is described by a combination of fixed and random terms:
\begin{equation}
Y_{i}(\bs,t) =  \mathbf{X}(\bs) \boldsymbol{\beta}_{z_k}(\bs) +\omega_i(\bs,t)+\lambda_i(\bs,t)+\epsilon_i(\bs,t)  \label{eq:mod}
\end{equation}
%
with $\epsilon_i(\bs,t)\stackrel{iid}{\sim} N(0, \sigma_{\epsilon,i}^2)$. Here $\mathbf{X}(\bs) = (1,X(\bs))$ and $X(\bs)$ is the elevation
of site $\bs$. The integer valued indicator $z_k(\bs) \subset{\mathbb{Z}^+}$ is the ecoregion label for the $k^{th}$ ecoregion tier: with
$k=1$ we have one ecoregion covering the entire country, while $k=5$ returns  the finer classification with 35 ecoregions. In general
$z_1(\bs) =1$, $z_2(\bs) \in \{1,2  \}$, $z_3(\bs) \in\{1,2 ,\dots 7 \}$, $z_4(\bs) \in\{1,2 ,\dots 13 \}$ and $z_5(\bs) \in\{1,2 ,\dots 35
\}$. Then $\boldsymbol{\beta}_{z_k}=(\beta_{0,z_k},\beta_{1,z_k})^\prime$ are regression coefficients, varying with the ecoregion.
%
The term $\lambda_i(\bs,t)$ describes the monthly effect of the annual cyclical behavior.
More precisely, we represent time on a circular scale with 1 year period. We assume that $h^*_t=h_t\mod L$ is a circular variable with
period $L=1$years, where $h_t=|t_l-t_d|$ is the temporal lag. This choice implies that e.g.: if $h_t= 1$years, then $h_t^*=h_t \mod 1$year
$=0$; if $h_t=1.1$years, then $h^*_t=0.1$ and so on. Then assuming $\lambda_i(\bs,t) \perp \lambda_i(\bs^\prime,t)$ for all $t$'s and
$i$'s, we describe the monthly effect of the annual cyclical behavior of each response by:
\begin{equation}\label{eq:circ}
\lambda_i(\bs,t) \sim N\left(0,\sigma^2_{cy,i} \exp( -\phi_{cy,i} h^*_t)\right),\;\; i=1,2,3
\end{equation}
where $\sigma^2_{cy,i},\phi_{cy,i}>0$ holds for both annual cyclical parameters.

Finally, the term $\omega_i(\bs,t)$ in (\ref{eq:mod}) is defined as a multivariate spatio-temporal Gaussian process (GP) with dependent
components. First we consider the multivariate GP $\mathbf{w}(\bs,t)=(w_1(\bs,t),w_2(\bs,t),w_3(\bs,t))^\prime$, where $w_i(\bs,t)\perp
w_j(\bs,t)$, for all $(\bs,t)$'s and $w_i\sim GP( \mathbf{0},C(h_{\bs},h_t; \boldsymbol{\theta}_i))$ where $C(h_{\bs},h_t;
\boldsymbol{\theta}_i)$ is a spatio-temporal correlation function and where $h_{\bs}=||\bs_l-\bs_q||$ and $h_t$ are the spatial and
temporal distances with $(h_{\bs},h_t)\in\mathbb{R}^2\times\mathbb{R}$. We choose the general non-separable space-time correlation
structure proposed by \cite{Gneiting2002}, defining $C(\cdot,\cdot;\cdot)$ as in his equation (14), i.e.:
\begin{equation}\label{eq:cov1}
 C(h_{\bs},h_t; \boldsymbol{\theta}_i) = \frac{1}{(\phi_{ti,i}|h_t|^{2\alpha_i}+1)^{
 \tau}}\exp\left(-\frac{\phi_{sp,i}\|h_{\bs}\|^{2\gamma_i}}{(\phi_{ti,i}|h_t|^{2\alpha_i}+1)^{\eta_i\gamma_i}}\right).
\end{equation}
%
Non-negative scaling parameters $\phi_{ti,i}$ and $\phi_{sp,i}$ are associated to time and space respectively, the smoothness parameters
$\alpha_i$ and $\gamma_i$ take values in $(0,1]$, the space-time interaction parameter $\eta_i$ ranges in $[0,1]$ and  $\tau \ge d/2$.
Following both \cite{Gneiting2002} and \cite{datta2016}, we set $\tau=1$, $\alpha=1$ and $\gamma=0.5$. Attractively, as $\eta_i$ decreases
towards zero, we achieve separability in space and time. Using the independent components of $\bw$, we can now define
$\boldsymbol{\omega}(\bs,t)=(\omega_1(\bs,t),\omega_2(\bs,t),\omega_3(\bs,t))^\prime$ as follows:
\begin{equation}\label{eq:coreg}
\boldsymbol{\omega}(\bs,t)=\mathbf{A}\mathbf{w}(\bs,t)\;
\end{equation}
where $\mathbf{A}\mathbf{A}^\prime=\mbox{Cov}\left(\omega_1,\omega_2,\omega_3 \right)=\bSigma$.
Now, letting $\mathbf{T}_i =  \ba_i  \ba_i^\prime$,
where $\mathbf{a}_i$ is the $i^{th}$ column of $\bA$, the covariance matrix for the process $\boldsymbol{\omega}$ at different times and
locations is given by:

\begin{equation} \label{eq:sigmald}
\bSigma_{l,q}= \mbox{Cov}( \boldsymbol{\omega}(\bs_l,t_l) ,  \boldsymbol{\omega}(\bs_q,t_q)  ) = \sum_{i=1}^3  \mathbf{T}_i C\left((||\bs_l-\bs_q||,|t_l-t_q|); \boldsymbol{\theta}_i\right).
\end{equation}
Remark that the choice of $\bA$ in equation \eqref{eq:coreg}  is not unique and has specific consequences on the process structure
\citep{Gelfand2004}, hence  a careful choice is required. A popular choice is the Cholesky decomposition of the symmetric matrix $\bSigma$
that produces a lower diagonal matrix. This decomposition induces an artificial ordering of the response variables in this setting, given
that the correlation structure of $\omega_1$ depends only on $C(\cdot,\cdot; \boldsymbol{\theta}_1)$, the one of $\omega_2$ depends  on
$C(\cdot,\cdot; \boldsymbol{\theta}_1)$ and $C(\cdot,\cdot; \boldsymbol{\theta}_2)$, while the correlation of $\omega_3$ depends  on
$C(\cdot,\cdot; \boldsymbol{\theta}_1)$, $C(\cdot,\cdot; \boldsymbol{\theta}_2)$ and $C(\cdot,\cdot; \boldsymbol{\theta}_3)$. To avoid this
artificial ordering, we propose to decompose by $\bSigma$ by a different approach: let $\boldsymbol{\Gamma} = diag(\gamma_1,\gamma_2,
\gamma_3)$ be the diagonal matrix of the square rooted eigenvalues of $\bSigma$ and $\boldsymbol{\Psi}$ be the orthogonal matrix of its
eigenvectors, such that $\boldsymbol{\Psi}^\prime\boldsymbol{\Psi}=\mathbf{I}$, we then let $\mathbf{A}=\boldsymbol{\Psi}
\boldsymbol{\Gamma}  \boldsymbol{\Psi}^\prime$.
%
{Such matrix $\mathbf{A}$ is symmetric by construction and its elements do not depend on the ordering of the eigenvalues.
Assume that $\mathbf{D}$ is a $3 \times 3$ matrix that changes the ordering of the elements of $\boldsymbol{\omega}(\mathbf{s},t)$. The
covariance matrix of $\mathbf{D}\boldsymbol{\omega}$ is then $\mathbf{D} \boldsymbol{\Sigma}\mathbf{D}'$ with eigenvectors as the columns
of matrix $\mathbf{D}\boldsymbol{\Psi}$. Now let $\mathbf{D} \boldsymbol{\Sigma}\mathbf{D}' = \mathbf{A}_*\mathbf{A}_*'$, then
$\mathbf{A}_* = \mathbf{D} \boldsymbol{\Psi} \boldsymbol{\Gamma}  \boldsymbol{\Psi}^\prime \mathbf{D}' = \mathbf{D}\mathbf{A}\mathbf{D}'$,
proving that $\mathbf{A}_* $ has the same values of $\mathbf{A}$  but arranged accordingly to the reordering matrix $\mathbf{D}$.}

\section{Implementation}\label{cyc:imp}
The huge dimension of the data, i.e. 360 spatial locations observed at 720 times, does not allow the implementation of a full multivariate
Gaussian process. This issue,  generally referred to as ``Big $n$ problem'' \citep{Jona2013b}, arises from the need to compute the
covariance matrix of the entire multivariate process, that in our case has dimension $3*720*360=777600$. Such a big matrix has to be stored
and inverted to compute the model likelihood, with a computational cost of the order of 
$O(777600^3)$. This  is not feasible even for very large computers or computer clusters.
For this reasons, an approximated approach has to be adopted. Several approaches to obtain computationally feasible approximations of
Gaussian processes have been proposed in the literature \citep[for example see][and references therein]{Jona2013b}. In this work we adopt
an efficient and accurate approximation of a Gaussian process recently proposed by \citet{datta2016}, namely the Nearest Neighbors Gaussian
Process (NNGP).

To address some issues related to the numerical stability of the estimation algorithm, we propose to rescale and standardize the response
 6.o,6ovariables. Only in the case of the monthly rain amount, in order to preserve the information about the zeroes, we simply rescale the
variable by its standard deviation. Hence, in what follows, $Y_1$ is rescaled while $Y_2$ and $Y_3$ are both standardized. Results are
presented according to the model (transformed) scale (except for the RMSE and figure \ref{fig:mappe}).

\subsection{NNGP}\label{cyc:nngp}
Letting $N$ be the number of observations in space and time, we denote their locations by $(\bs_n,t_n)$, $n=1,\dots , N$ and we let
$\boldsymbol{\omega}_n=(\omega_1(\bs_n,t_n),\omega_2(\bs_n,t_n),\omega_3(\bs_n,t_n))^\prime$ with $\boldsymbol{\omega}=
(\boldsymbol{\omega}_1, \dots , \boldsymbol{\omega}_N)^\prime$.  If $f(\cdot)$ is a generic density function, then the joint distribution
of the whole set of observations is given by
\begin{equation}\label{eq:jointomega}
f(\boldsymbol{\omega}) = \prod_{n=1}^N f(\boldsymbol{\omega}_n| \boldsymbol{\omega}_{n-1},\dots, \boldsymbol{\omega}_1)
\end{equation}
with $\boldsymbol{\omega}_0=\emptyset$.
In \eqref{eq:jointomega} $f(\boldsymbol{\omega})$  and $f(\boldsymbol{\omega}_n| \boldsymbol{\omega}_{n-1},\dots, \boldsymbol{\omega}_1)$
are Gaussian densities of size $3\cdot N$ and $3$, respectively. Notice that, though there is no univocal definition of a space-time
ordering of observed locations, \eqref{eq:jointomega} is a valid representation of the joint density for any given ordering.

Let $\boldsymbol{\Omega}_n = (\boldsymbol{\omega}_{n-1},\dots, \boldsymbol{\omega}_1)'$ be the conditional set of $\boldsymbol{\omega}_n$
in \eqref{eq:jointomega}  and let $\boldsymbol{\Omega}_n(m) \subseteq  \boldsymbol{\Omega}_n $ be a set that contains at most $m$ elements
of $\boldsymbol{\Omega}_n $. With the NNGP the joint distribution of the whole set of observations in \eqref{eq:jointomega} is by
\begin{equation} \label{eq:ref} \prod_{n=1}^N f(\boldsymbol{\omega}_n| \boldsymbol{\Omega}_n(m)). <
\end{equation}
Indeed the quality of the approximation increases with $m$ and, as shown by \cite{datta2016}, and the best results are achieved if we chose
the $m$ elements of $\boldsymbol{\Omega}_n$ that have the higher correlation with $\boldsymbol{\omega}_n$.

 To  implement the NNGP three decisions have to be made:
 \begin{itemize}
    \item how to order the observations;
    \item how to choose the the value of $m$;
    \item how to choose the elements of $\boldsymbol{\Omega}_n(m)\subseteq\boldsymbol{\Omega}_n$.
 \end{itemize}

\paragraph{The ordering}
A natural ordering is immediately available for the time dimension, but there is not a unique way to order observations in space at a given
time. The way we order spatial locations has a strong influence on the definition of how candidate locations enter
$\boldsymbol{\Omega}_n(m)$. Here we follow \citet{datta2016} and order locations first according to one of the two coordinates and then
according to the other. This ensures that $\boldsymbol{\Omega}_n$ includes observations spatially and temporally close to
$\boldsymbol{\omega}_n$.

\paragraph{The value of $m$}
Compared to the size of the problem, the number of neighbors $m$ should be small in order to obtain a computational gain.
\citet{datta2016a} showed that, assuming that the elements of  $\boldsymbol{\Omega}_n(m)$ are ``close enough'' (correlated or
geographically close) to    $\boldsymbol{\omega}_n$, $m \in \{10,\dots , 20\}$ produces an approximation almost indistinguishable from the
original process.

\paragraph{The elements in $\boldsymbol{\Omega}_n(m)$}
Again, following \citet{datta2016a}, the best choice for $\boldsymbol{\Omega}_n(m)$ is to take the $m$ elements that have higher
correlation with  $\boldsymbol{\omega}_n$. In a purely univariate temporal or spatial setting, assuming that the correlation decreases with
the distance, the optimal choice for the elements $\boldsymbol{\Omega}_n(m)$ would consider observations spatially/temporally closer to
$\boldsymbol{\omega}_n$.
In a spatio-temporal setting with non-separable correlation function,  there is not a one to one relation between distances and
correlation, since a spatio-temporal distance is not uniquely defined. In an univariate spatio-temporal setting, \citet{datta2016a} propose
an adaptive approach in which $\boldsymbol{\Omega}_n(m)$ is defined at each MCMC iteration as the set that has the higher correlation with
$\boldsymbol{\omega}_n$.

Basing the choice of $\boldsymbol{\Omega}_n(m)$ on correlations would imply to consider all possible sets of $m$ neighbors at each point
for each MCMC iteration. In this work we prefer not to follow this approach, mostly for computational reasons as, unlike in
\citet{datta2016a}, here we deal with a very large multivariate spatio-temporal data base.
Hence we propose
to define a  spatio-temporal distance as shown in expression \eqref{eq:distadjust} and to include in $\boldsymbol{\Omega}_n(m)$ the
locations with smaller distances from  $\boldsymbol{\omega}_n$. Obviously, the spatial and temporal dimensions have different scales, so we
adjust the spatio-temporal distance euristically as follows:
\begin{equation}\label{eq:distadjust}
\sqrt{\left(h_{s}\frac{2}{30}\right)^2+h_t^2},
\end{equation}
assuming that one year has the same weigh as $ 150$Km's. Our choice is justified by the following considerations: in each neighborhood we
want to include information on the spatial dependence, the time dependence and the cross-correlation structure, furthermore we need
information on the annual cyclical component. Equation \eqref{eq:distadjust} ensures that  the generic point $(\bs_n,t_n)$ has
approximately $m$ neighbors, $\sqrt{m}$ of which are observed at the same time and at different locations,  $\sqrt{m}$  share the same
spatial location and are observed at different times and the remaining are observed at different times and locations. Furthermore, in order
to learn about the annual cyclical component, we may have to modify the points at the boundaries of $\boldsymbol{\Omega}_n(m)$. This is
done in such a way that, for example, the neighborhood of location $\bs$-January 2000 includes location $\bs$-January 1999 and
$\bs$-February 1999.

\subsection{Implementation details}
In our setting $E(\boldsymbol{\omega})= \mathbf{0}_N$,  then using standard results from the multivariate normal theory,  we can write
 \begin{equation}
 f(\boldsymbol{\omega}_n| \boldsymbol{\Omega}_n(m) ) = \phi_3(\boldsymbol{\omega}_n| \mathbf{B}_n \boldsymbol{\Omega}_n(m) , \mathbf{F}_n)
 \end{equation}
where $\phi_3(\boldsymbol{\omega}_n| \mathbf{B}_n \boldsymbol{\Omega}_n(m), \mathbf{F}_n)$ is the $3-$variate normal distribution with mean
$\mathbf{B}_n \boldsymbol{\Omega}_n(m) $ and covariance matrix $\mathbf{F}_n$. Parameters $\mathbf{B}_n$ and $\mathbf{F}_n$  depend on the
Gneiting correlation function parameters, on $\boldsymbol{\Sigma}$ and on the distances between the spatio-temporal  locations in
$(\boldsymbol{\omega}_n,\boldsymbol{\Omega}_n(m) )$.

Our data are observed over 360 spatial locations, that are the same at each time point. Given the set of $m$ values we explored
($m=10,15,20$), the maximum temporal distance between $\boldsymbol{\omega}_n$ and the elements of  $\boldsymbol{\Omega}_n(m)$ is equal to 1
year, with the exception of the first 12 months in the database that have a maximum distance of less then one year. Hence, starting from
the 13$^{th}$ time-point onwards, the parameters $\mathbf{B}_n$ and $\mathbf{F}_n$ at the same spatial location will be the same, as  they
are based on the same distance matrices. Then we only need to compute $\mathbf{B}_n$ and $\mathbf{F}_n$ for the first 13 times and 360
spatial locations, thus obtaining a huge computational gain. 
Notice that in this setting the computation of the full conditionals of $\omega_n$ implies only a time window of two years at each sampled
location. Given that after the first 12 months the distances between points start to repeat, we need to compute the full conditionals for
the first 13 months and the last 12 months, as in these latter cases the two years time window is not available.

\paragraph{Grid prediction}
Let $\mathbf{Y}_0 = (\mathbf{Y}(s_{N+1},t_{N+1}), \dots \mathbf{Y}(s_{N+720},t_{N+720}))'$
be the 3-variate time series of $Y$'s at a spatial grid point 
and let define $\boldsymbol{\Omega}_{y,N+j} = ( \mathbf{y}_{N+j-1}, \dots ,  \mathbf{y}_{1})$ as the conditioning set of
$\mathbf{y}_{N+j}$, with $j \in \{1,\dots , 720\}$ and $ \mathbf{y}_{n} =
(y_1(\mathbf{s}_n,t_n),y_2(\mathbf{s}_n,t_n),y_3(\mathbf{s}_n,t_n))'$.
In like vein, we define $\Omega_{y,N+j}(m)$  as the set of $m$ nearest neighbors of $\mathbf{y}_{N+j}$, based on the distance \eqref{eq:distadjust}.
Notice that, since $\Omega_{y,N+j}$  contains all spatio-temporal locations, the set of
$m$ nearest neighbors $\Omega_{y,N+j}(m)$  can contain temporal indexes that are even higher than $t_{N+j}$, e.g. in the set of neighbors
of the first time of a grid point, there can be points in the second or third time.

We want to obtain samples of $\mathbf{Y}_0 $ from the predictive density
\begin{equation} \label{post}
f(\mathbf{y}_{0}|  \mathbf{y}^{O} ) = \int  \prod_{j=1}^{720}  f(\mathbf{y}_{N+j}|\boldsymbol{\Omega}_{y,N+j},
\boldsymbol{\theta})f( \mathbf{y}^{M} , \boldsymbol{\theta}|\mathbf{y}^{O}) d\boldsymbol{\theta} d\mathbf{y}^{M}.
\end{equation}
where  $\mathbf{y}^{M}$ and  $\mathbf{y}^{O}$ are subsets of $\mathbf{y}=(\mathbf{y}_1,\dots ,\mathbf{y}_N)'$ composed of, respectively,
missing and observed  data, $\boldsymbol{\theta}$ contains all model parameters and $f( \mathbf{y}^{M} ,
\boldsymbol{\theta}|\mathbf{y}^{O})$ is the posterior distribution. Our interest is also in the prediction of the annual cyclical component
$\boldsymbol{\lambda}_0 = (\boldsymbol{\lambda}_{N+1},\dots, \boldsymbol{\lambda}_{N+12})$, where  $\boldsymbol{\lambda}_{N+j} =
(\lambda_1(s_j,t_j),\lambda_2(s_j,t_j),\lambda_3(s_j,t_j))'$. 
We then sample from the following predictive density:
 \begin{equation} \label{postseas}
 f(\boldsymbol{\lambda}_0| \mathbf{y}^{O} ) = \int  f(\boldsymbol{\lambda}_0| \mathbf{y}_{0},  \mathbf{y},  \boldsymbol{\theta} )  \prod_{j=1}^{720}  f(\mathbf{y}_{N+j}|\boldsymbol{\Omega}_{y,N+j},
 \boldsymbol{\theta})f( \mathbf{y}^{M} , \boldsymbol{\theta}|\mathbf{y}^{O}) d\boldsymbol{\theta} d\mathbf{y}^{M} d \mathbf{y}_{0}.
 \end{equation}
The density $ f(\mathbf{y}_{N+j}|\boldsymbol{\Omega}_{y,N+j}, \boldsymbol{\theta})$ in \eqref{post} and \eqref{postseas} is a trivariate
normal, but as with equation \eqref{eq:jointomega}, estimation of its parameters requires the computation/inversion of a covariance matrix
of dimension $N+j$ \citep{Banerjee2014}.
 We then use the NNGP  to approximate the predictive densities and substitute $\boldsymbol{\Omega}_{y,N+j}(m)$ to $\boldsymbol{\Omega}_{y,N+j}$ in both expressions. After model
fitting, posterior samples from \eqref{post} and \eqref{postseas} can be obtained using standard Monte Carlo procedures.

\section{Results and discussion}\label{cyc:result}
We estimated nine different models, varying the number of neighbors in the NNGP, $m=\{10,15,20\}$, and the ecoregional hierarchical tier,
$k \in \{3,4,5\}$. The MCMC was implemented with 100000 iterations, a burn-in phase of 70000 and thinning by 12, keeping 2500 samples for
posterior inferences. Posteriors estimates were obtained in about three days of a fast computer cluster, as specified below.  Model choice
was performed using the DIC \citep{spiegelhalter2002} and results are reported in table \ref{tab:dic}. As expected, the largest number of
neighbors always returns the smallest DIC value for a given $k$. In bold we highlight the ``best'' model that suggest to aggregate
ecoregions into 7 distinct provinces (see figure \ref{fig:ecoregionslev}). Provinces represent the highest and most general ecoregional
tier among those considered with the model implementation. Therefore, this result is consistent with a principle widely adopted by
hierarchical approaches for the ecological classification of land. This basic principle states that climate acts as a primary environmental
factor in determining the broad-scale ecosystem variation. On the contrary, factors such as geomorphology and soil features assume an equal
or greater importance than climate only at lower levels \citep{Bailey2004,MUCHER2010}.

\begin{table}[t]
    \centering
    \begin{tabular}{cc|ccc}
        \hline
        &&&m&\\ \hline \hline
        &&10&15&\textbf{20}\\ \hline  \hline
        &\textbf{3}& 4593496&3434861&\textbf{3362126}\\
        k &4&4797839 &4518074&3853829 \\
        &5& 6793930&5537513&5432627 \\  \hline \hline
    \end{tabular}
    \caption{Model choice, DIC values for different choices of the hierarchical ecoregional tier ($k$) and neighborhood size ($m$) in the NNGP approximation.} \label{tab:dic}
\end{table}

\begin{figure}[t] 
\centering
\includegraphics[scale=0.5]{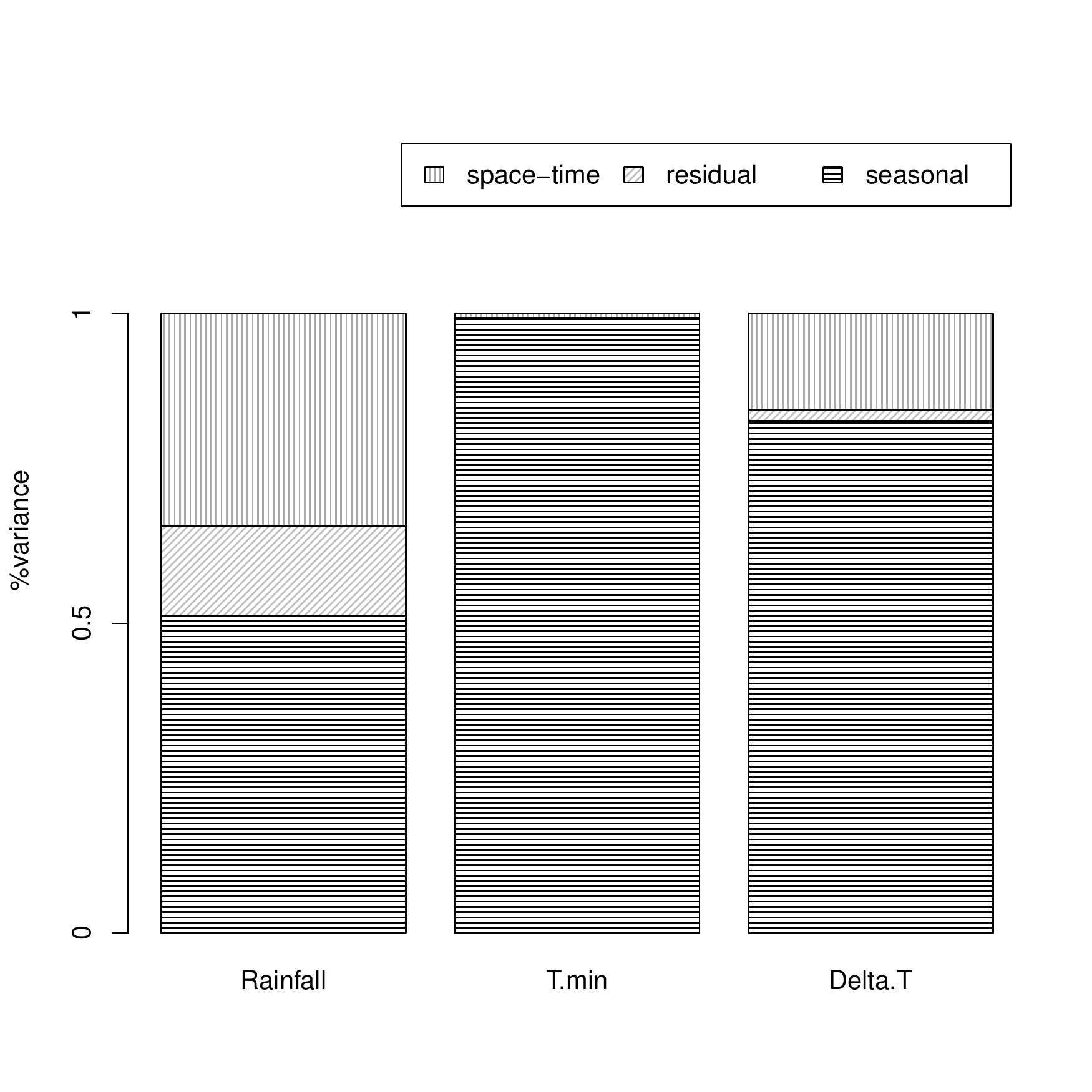}
\caption{Proportion of  the space-time, seasonal and residual components of the variance for each climate variable}\label{fig:percvar}
\end{figure}

{Posterior estimates of the Gaussian Process parameters (see equations \eqref{eq:circ} and \eqref{eq:cov1}) and their variances  are
reported in table \ref{tab:posteriorpar}, while in figure \ref{fig:percvar} the proportion of the variance of the seasonal, space-time and residual term over their sum is reported, this in order to descripe the relevance of each term in explaining the totalvariation of each variable. It is worth noticing that for the minimum temperature almost the entire variation can be ascribed to the seasonal component, while for the thermal excursion a 15.5\% is due to the spatiotemporal term and a negligible contribution comes from the residual part. The precipitation has a different behaviour, a large portion of variation is seasonal (51\%), but now the space-time dynamic has more relevant role, while 14.6\% of variation is left unexplained. These behaviour of the three variables is perfectly compatible with the physics of the phenomena described, where rainfall is more influenced by local events here not available.} 
\begin{table}[t]
        \centering 
 {\tiny    \begin{tabular}{cc|cccc}
        \hline
        && $\phi_{sp}$ & $\phi_{ti}$& $\phi_{cy}$&  $\eta$\\
        \hline\hline
        \multirow{2}{*}{$Y_1$} & Est.        & 0.188  &28.979 &15.210  & 0.774  \\
        & (CI) & (0.184 0.192)& (28.871 29.072)& (14.972 15.394)& (0.774 0.775) \\
        \multirow{2}{*}{$Y_2$} & Est.        &0.138   & 9.628 & 10.176 &0.943  \\
        & (CI) & (0.137 0.140)& (9.476 9.750)& (10.102 10.239)& (0.942 0.943) \\
        \multirow{2}{*}{$Y_3$} & Est.        & 0.431  &23.814  & 9.760  &  0.166     \\
        & (CI) & (0.429 0.432)& (23.576 23.995)& (9.690 9.835)& (0.165 0.168)\\
        \hline\hline
        && $\sigma^2_{cy}$& $\sigma^2_{\epsilon}$&$\sigma^2_{\omega}$\\
        \hline\hline
        \multirow{2}{*}{$Y_1$} & Est.        &0.617 &   0.176  & 0.413   \\
        & (CI) &   (0.612 0.623)& (0.175 0.178)&0.409 0.416) \\
        \multirow{2}{*}{$Y_2$} & Est.        &  6.968&      0.008 & 0.050 \\
        & (CI) &  (6.830 7.092)& (0.008 0.008)& (0.050 0.051) \\
        \multirow{2}{*}{$Y_3$} & Est.        &   2.799&0.062  &0.525         \\
        & (CI) &  (2.647 2.919)& (0.061 0.062 )&(0.519 0.532)\\
        \hline\hline
    \end{tabular}
}    \caption{Posterior estimates of the Gaussian process and annual cyclical component parameters, as in equations \eqref{eq:cov1} and
\eqref{eq:circ}, respectively.}\label{tab:posteriorpar}
\end{table}
Table \ref{tab:posteriorpar} shows a clear evidence that the three components are non-separable in space and time, as the $\eta$ parameter
is never close to 0. The practical ranges  and covariances of the three components provide useful information on the extent of the spatial,
temporal and annual cyclical dependence. The spatial practical ranges of the process components are 15.95km ($Y_1$), 21.676km ($Y_2$) and
6.967km ($Y_3$), suggesting a much more localized behavior of the temperature range with respect to the other two variables. In terms of
time dependence, we have the following practical ranges: 37.78 days ($Y_1$), 113.73 days ($Y_2$) and 45.98 days ($Y_3$). These values
highlight the length of time windows of correlated behavior for each component, suggesting a similar temporal dependence for the rainfall
and the temperature range. Finally the annual cyclical effect $\phi_{\mbox{cy}}$ (practical ranges: 71.99 days ($Y_1$), 107.60 days
($Y_2$), 112.19 days ($Y_3$)) highlights a similar behavior for the second and third variables, as expected: a shorter cycle is estimated
for the rain, while annual cycles are longer and almost seasonal (4 months long) for both temperatures. Correlation between climate
variables are also obtained and they are all significantly different from zero. Rain and minimum temperature are positively correlated
(0.210 with a narrow 95$\%$ credible interval (0.208,0.213)), while rain an temperature range are negatively correlated (-0.214 with again
a narrow 95$\%$ CI (-0.218,-0.211)). The minimum temperature and the temperature range are negatively correlated, showing a  stronger
relation, as expected (-0493, 95$\%$ CI (-0.499,-0.497)). This proposal allows to gather considerably more information on the joint
behavior of the climate variables than previous studies where an older version of the database was also used to correlate climate data with
altitude \citep{Blasi2007,Blasi2007a}.

\begin{figure}[t]
\centering
\subfloat[Rainfall January]{\includegraphics[width=4cm, height=4cm]{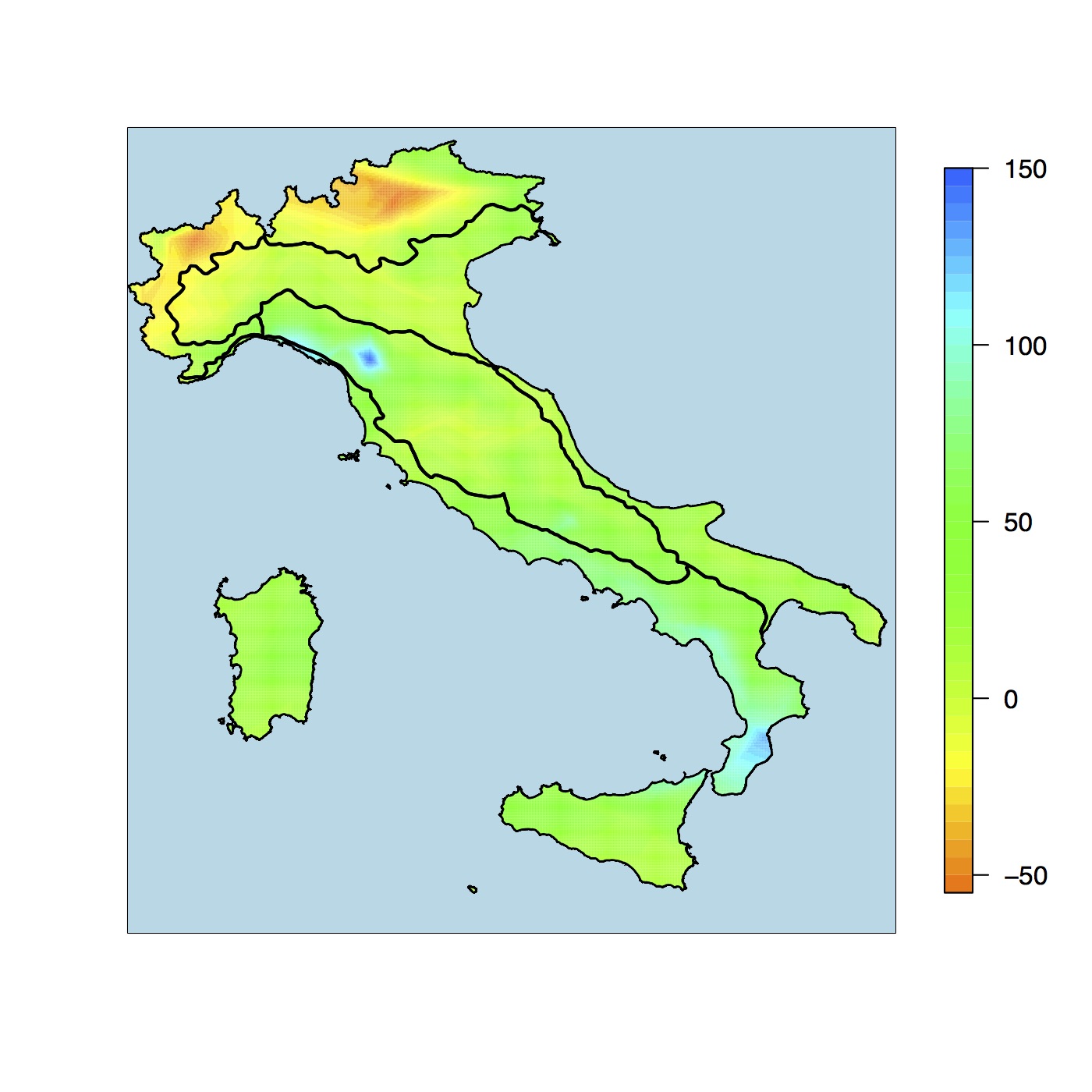}}
\subfloat[Temperature min. January]{\includegraphics[width=4cm, height=4cm]{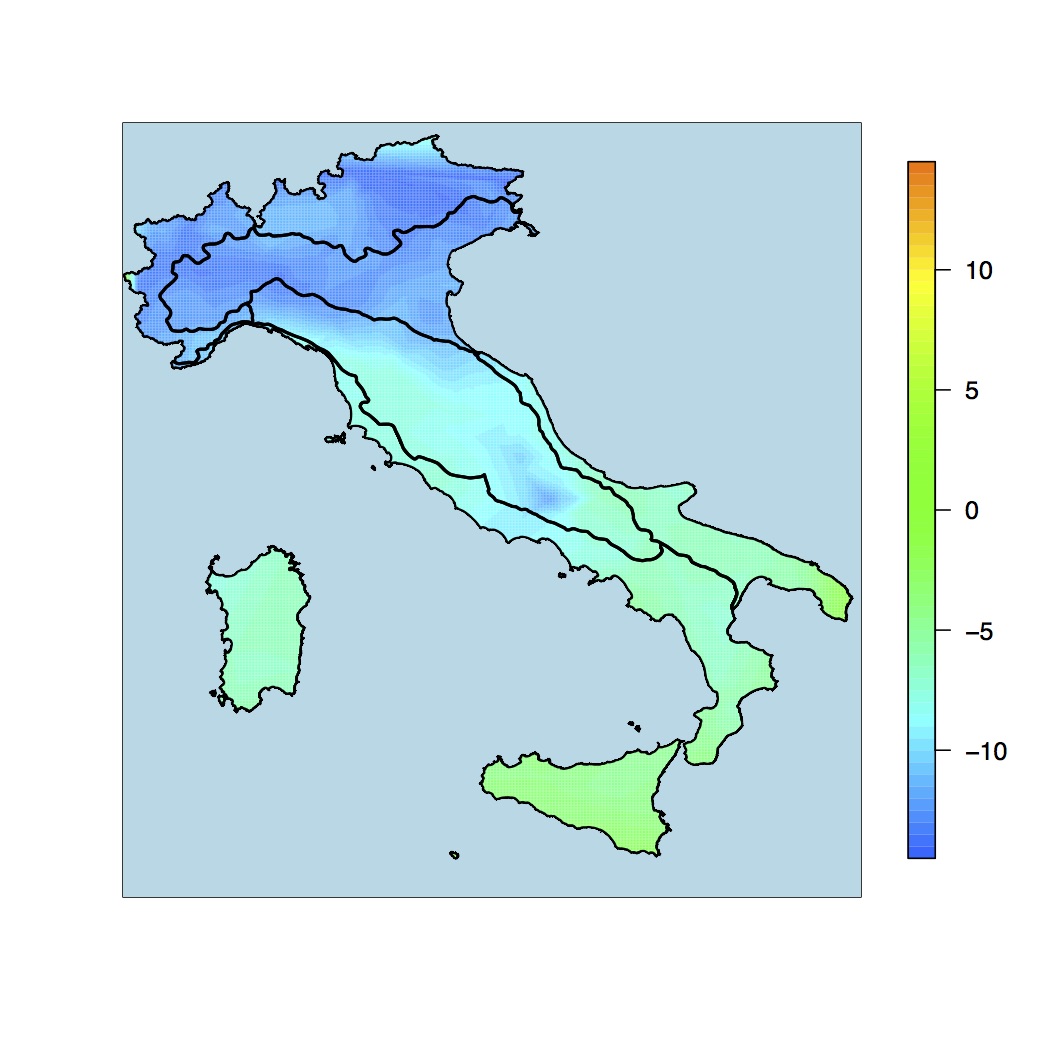}}
\subfloat[$\Delta$ Temperature January]{\includegraphics[width=4cm, height=4cm]{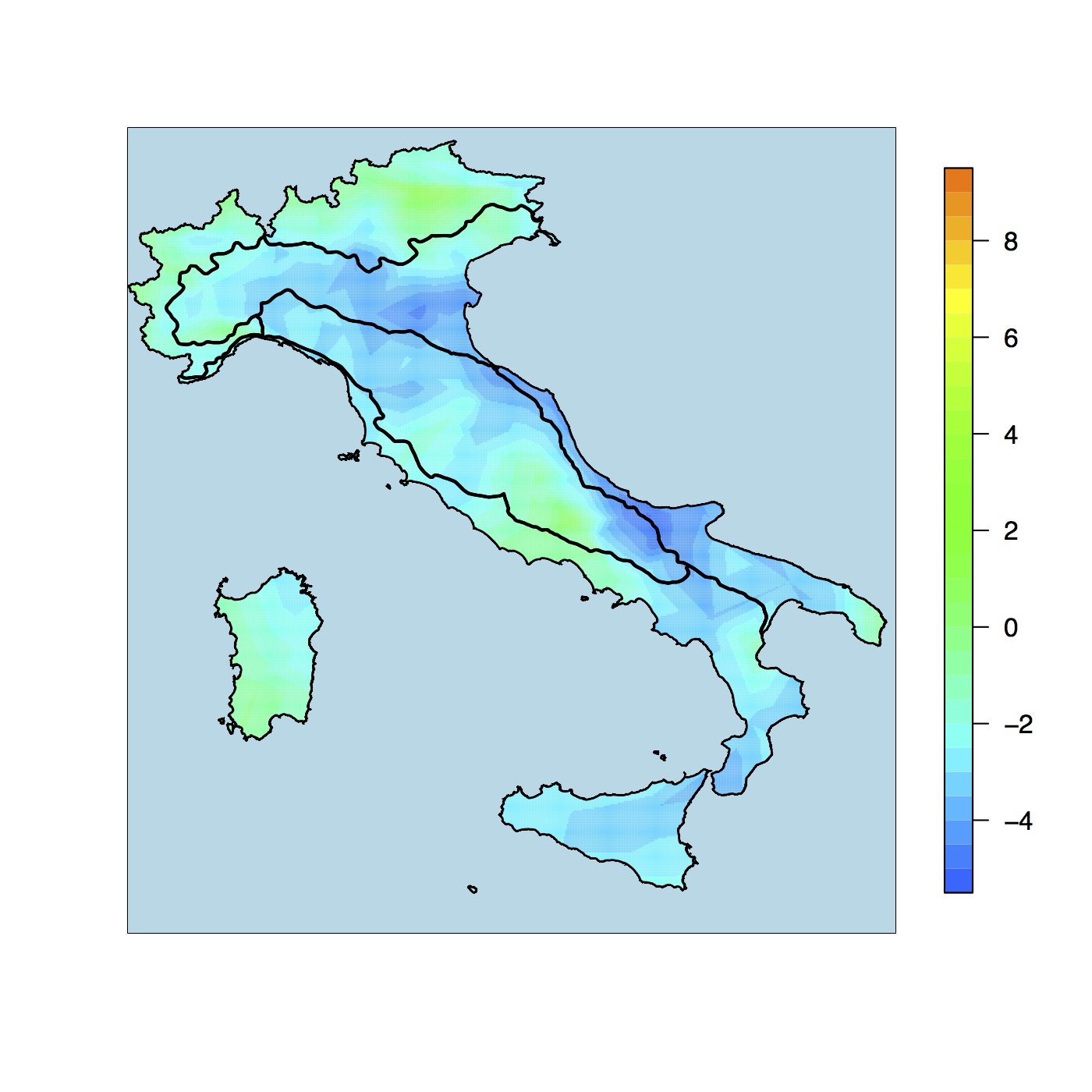}}\\
\subfloat[Rainfall August]{\includegraphics[width=4cm, height=4cm]{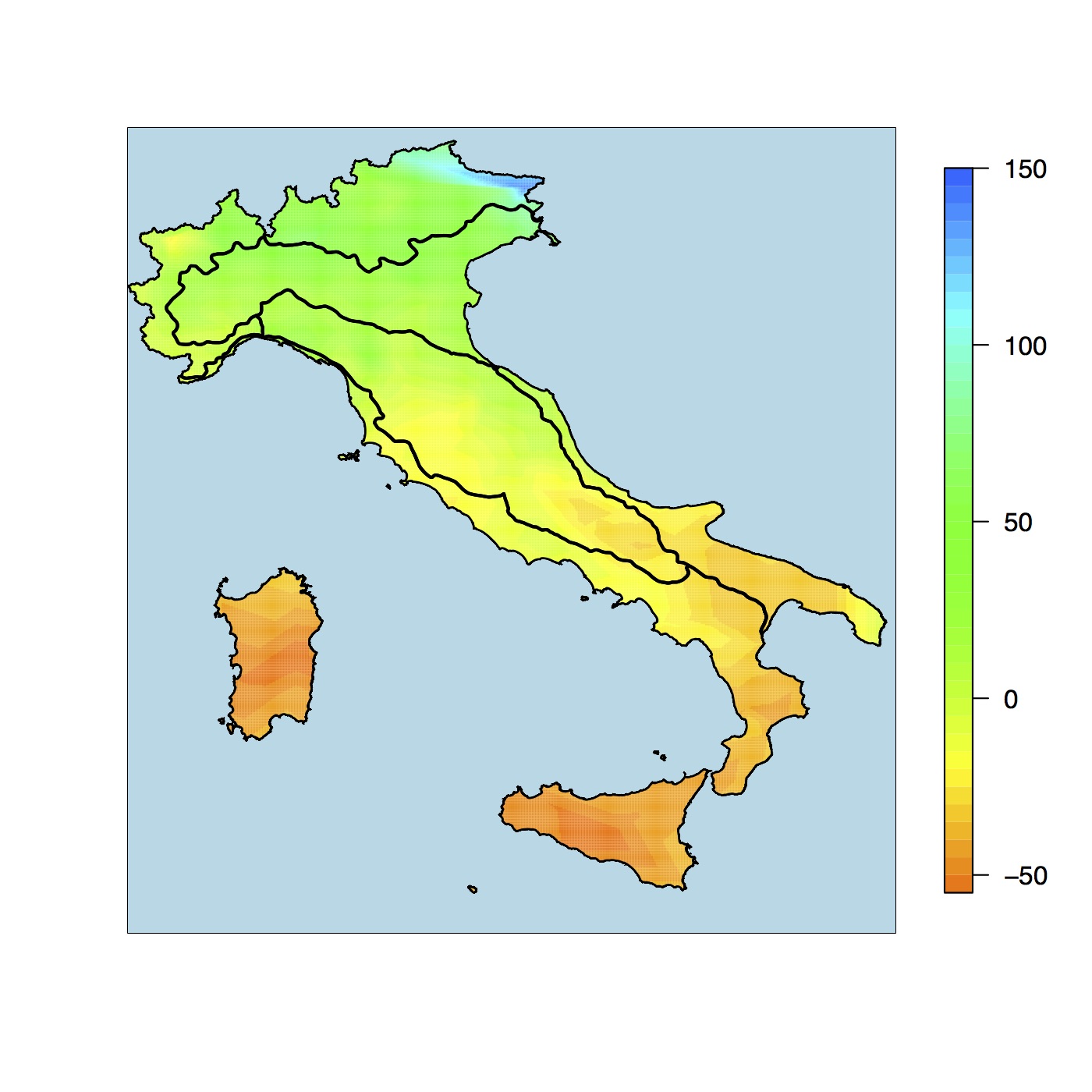}} \subfloat[Temperature min.
August]{\includegraphics[width=4cm, height=4cm]{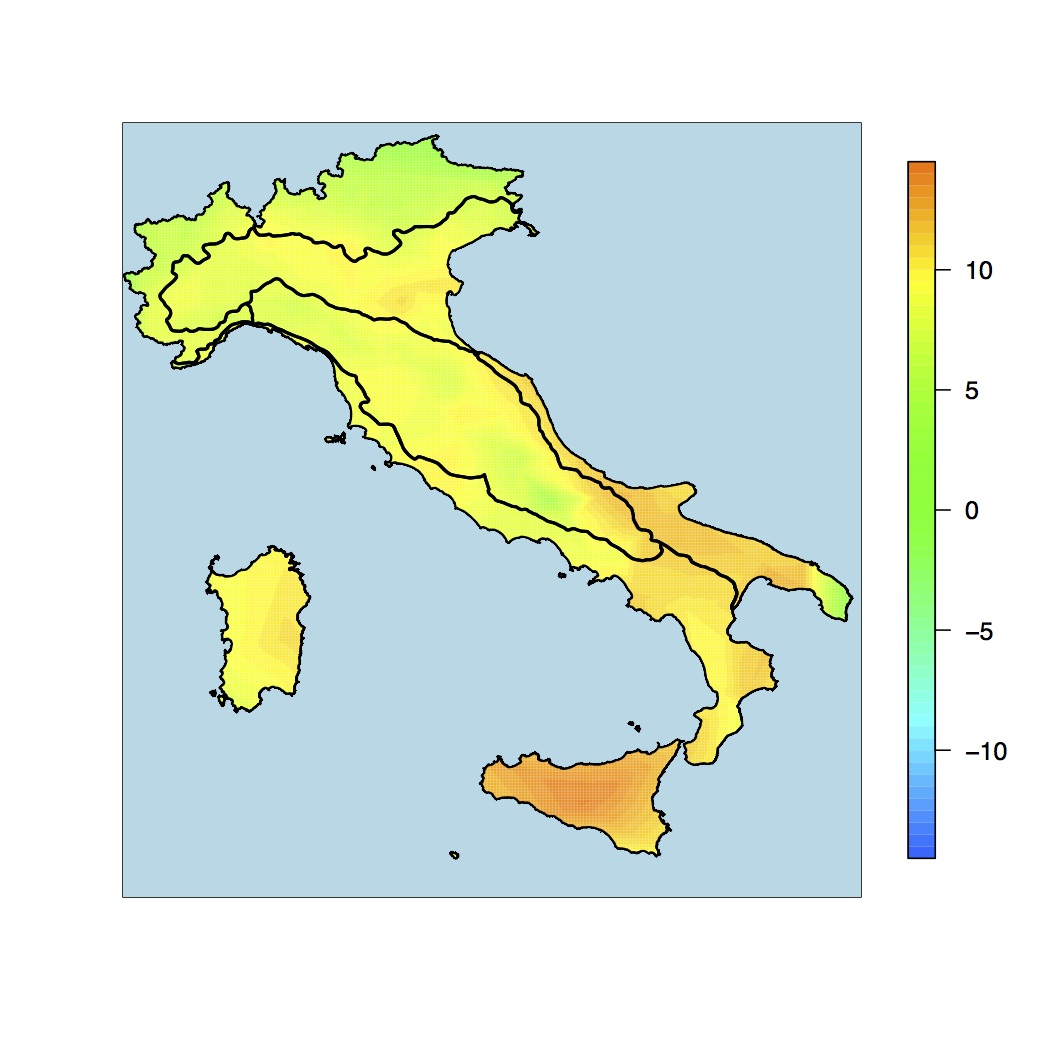}} \subfloat[$\Delta$ Temperature
August]{\includegraphics[width=4cm, height=4cm]{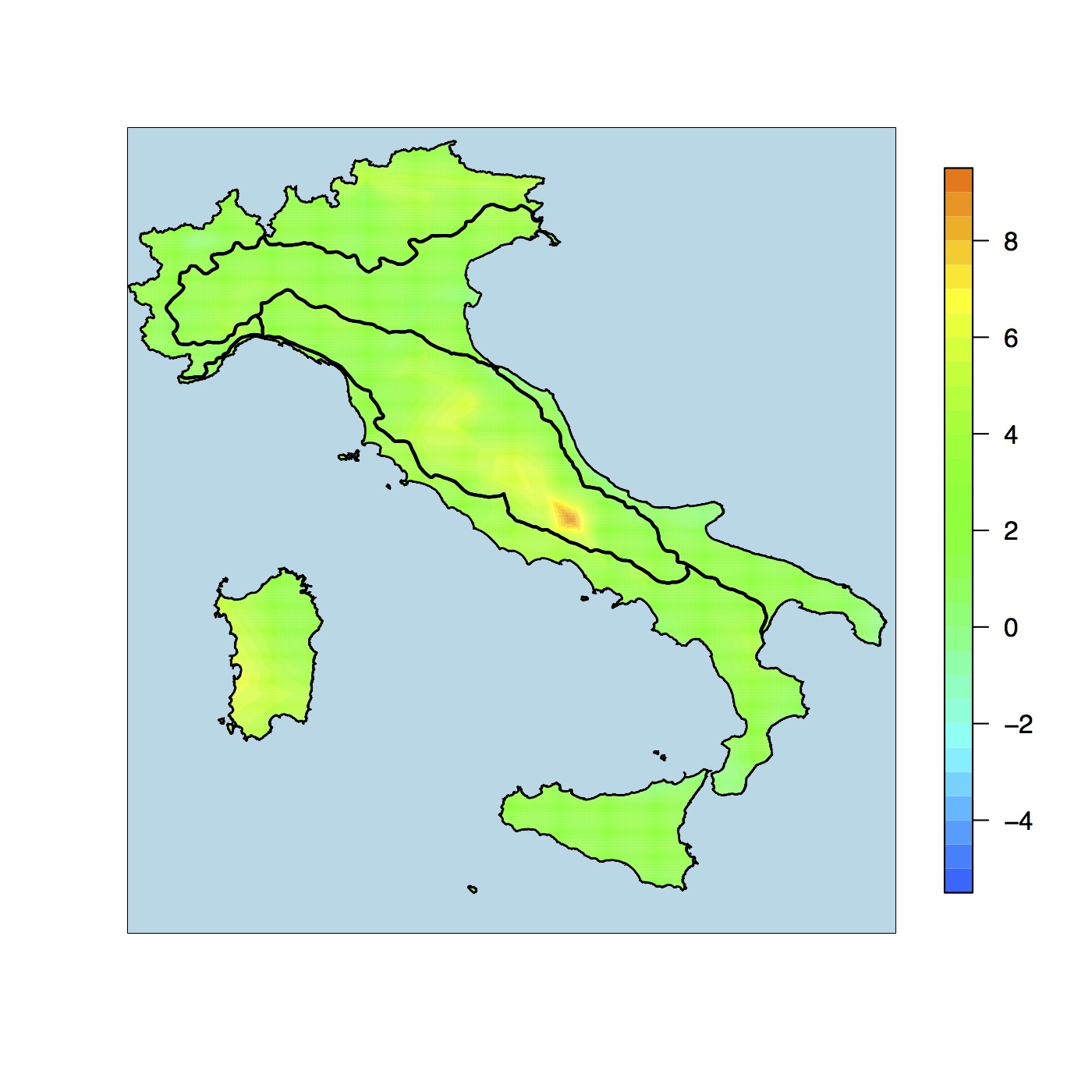}} \caption{Maps of the monthly effects of January (a,b,c) and August
(d,e,f) on the annual cycles of the three components of the process.}\label{fig:mappe}
\end{figure}

In figure  \ref{fig:mappe} examples of  maps of two monthly effects of the annual cyclical components are reported. The maps for the months
of January and August have been chosen as representative of the factors affecting the composition of ecosystems and their distribution over
the Italian territory. Above all, these factors include moisture availability in the different seasons, winter cold and summer drought.
Firstly, the model was able to show some interesting seasonal patterns of rainfall (fig. \ref{fig:mappe}a and \ref{fig:mappe}d). These
include: (i) the continental regime of the Alpine Province, the only  region with larger rainfall values in summer than in winter months;
(ii) the transitional character of the Po Plain Province towards a more Mediterranean regime, with lower summer rainfall; (iii) the very
clear latitudinal gradient in both the Apennine and peninsular Tyrrhenian Provinces, which mainly reflects the varying distance from the
coast of the mountain reliefs. More local patterns are suggested as well, that however need a more in depth investigation at the Section
and/or Subsection ecoregional levels. These include, for example, the longitudinal summer gradient in rainfall between Eastern and Western
Alps and the marked summer rainfall decrease in some Southern peninsular and main island sectors.
Secondly, winter cold (fig. \ref{fig:mappe}b) clearly characterises both the Alpine and Po Plain Provinces within the Temperate Division.
On the contrary, the variable behaviour within the Apennine Province should be further investigated in order to highlight in detail the
differences with the Tyrrhenian Province of the Mediterranean Region. Patterns that need to be characterised at lower ecoregional levels
emerged in this case as well. These include, for example, the latitudinal gradient along the Adriatic Province and the differences between
the two main Tyrrhenian islands.
The third component of the process, among the several features, highlights the relevance of reduced winter temperatures and their variation in characterising the termic continentality of the Po Plain and Adriatic Provinces. It also confirms that higher temperatures occur in both the Tyrrhenian and the Adriatic Mediterranean Provinces.

In tables \ref{tab:ecobeta1} and \ref{tab:ecobeta2} the posterior estimates of the intercepts and regression coefficients distinguished by
ecoregion are reported. Remark that all estimates for the 4th ecoregion (1D) are not relevant and very different from the other values, due
to the presence of only one monitoring station in the given area. This suggests to consider the aggregation of the 4th ecoregion to one of
its neighbors for future  investigations, as recently tested in  the first report on the Italian natural capital\footnote{{\tiny Italian Natural
Capital Committee (INCC), 2017. 1st Report on the State of Natural Capital in Italy (synthesis). Available at:
\url{http://www.minambiente.it/sites/default/files/archivio/allegati/sviluppo_sostenibile/sintesi_raccomandazioni_primo_rapporto_capitale_naturale_english_version.pdf}}
}.
Estimates of the model intercepts $\beta_{0}$'s allow to analyse the behaviour of each component in the specific ecoregion. The only
estimate that shows a value close to zero is for temperature range in ecoregion 1B, the Po Plain Province, suggesting a very small
temperature range. All ecoregions are well characterized with some overlapping of credible intervals for each variable, suggesting
similarities between areas. Similar behavior in terms of rainfall ($Y_1$) can be found in ecoregions 1B, 2A, 2B and 2C ($j=2,5,6,7$), while
1A, 1B and 2C ($j=1,2,7$) show similarities in terms of minimum temperature ($Y_2$) and only 1A and 2C ($j=1,7$) show intervals estimates
overlapping for the temperature range ($Y_3$). Notice that 1B covering the Po Plain is a very variable area where a transition from the continental to the mediterranean behaviours occurs. %
The relation with the elevation described by the estimates of the regression coefficients $\beta_{1}$'s often admits the zero value in the
95$\%$ credible interval. This is likely linked to the presence of a latitudinal gradient in the area. For example, in the Po Plain Province
(1B) a large area is divided by the Po river in a northern sector with continental regime and a southern sector with Apennine regime, as already highlighted for the effects of the cyclical components. The
Alpine province (1A) is associated to regression coefficients that are all quite far from zero and this can be linked to the absence of a
latitudinal gradient, being the region affected only by a longitudinal variation. Moreover the area is characterized by a considerable relief energy (large quota gradient). 

\begin{table}[t]
        \tiny
    \centering
    \begin{tabular}{c|c|cccc}
            \hline
    && $\beta_{0,1}$ & $\beta_{0,2}$ & $\beta_{0,3}$ & $\beta_{0,4}$ \\
        \hline  \hline
    \multirow{2}{*}{$Y_1$} & Est.        & 1.348 & 0.741&0.965&-0.050  \\
        & (CI) & (1.280 1.400)& (0.668 0.794)& (0.924 1.003)& (-8.155  7.777) \\
    \multirow{2}{*}{$Y_2$} & Est.        &0.207 &0.225 &0.094&-0.005    \\
& (CI) & (0.189 0.234)& (0.207 0.240)& (0.084 0.106)& (-1.746  1.758) \\
    \multirow{2}{*}{$Y_3$} & Est.        &0.155  & 0.008&0.779 &0.083    \\
& (CI) & (0.055 0.201 )& (-0.069  0.097)& (0.737 0.844)& (-4.919  4.748)\\
            \hline          \hline
    &&  $\beta_{0,5}$ & $\beta_{0,6}$ & $\beta_{0,7}$\\
        \hline  \hline
    \multirow{2}{*}{$Y_1$} & Est.        & 0.633&0.740 &0.783         \\
        & (CI) &  (0.520 0.787)& (0.710 0.774)& (0.745 0.838) \\
    \multirow{2}{*}{$Y_2$} & Est.        &0.878&0.457&     0.277      \\
& (CI) & (0.838 0.910)& (0.445 0.464)& (0.263 0.295) \\
    \multirow{2}{*}{$Y_3$} & Est.        & -1.965& -0.070&   0.196       \\
& (CI) &  (-2.038 -1.856)& (-0.125 -0.038)& (0.127 0.264) \\
            \hline          \hline
    \end{tabular}\caption{Point estimates of the intercepts at each ecoregion and for each Gaussian process component. Ecoregions are coded as follows $1=1A$,
$2=1B$, $3=1C$, $4=1D$, $5=2A$, $6=2B$ and $7=2C$.}\label{tab:ecobeta1}
\end{table}

\begin{table}[t]
    \tiny
    \centering
    \begin{tabular}{c|c|cccc}
        \hline
        && $\beta_{1,1}$ & $\beta_{1,2}$ & $\beta_{1,3}$ & $\beta_{1,4}$\\
        \hline  \hline
        \multirow{2}{*}{$Y_1$} & Est.        & 0.029&0.498&0.248&63.022      \\
        & (CI) & (-0.022  0.068)& (0.189 0.742)& (0.208 0.294)& (-650.237  788.956) \\
        \multirow{2}{*}{$Y_2$} & Est.        & -0.747 & -0.322&-0.452 &73.423 \\
        & (CI) & (-0.761 -0.729)& (-0.414 -0.242)& (-0.470 -0.437)& (-86.057 231.301) \\
        \multirow{2}{*}{$Y_3$} & Est.        &  -0.366& -0.094 & -1.073&-117.311   \\
        & (CI) & (-0.428 -0.305)& (-0.648  0.423)& (-1.177 -1.023)& (-544.667  338.024) \\
        \hline          \hline
&&  $\beta_{1,5}$ & $\beta_{1,6}$ & $\beta_{1,7}$\\
        \hline  \hline
        \multirow{2}{*}{$Y_1$} & Est.        & -0.532&0.548&0.416          \\
        & (CI) &   (-1.250  0.241)& (0.502 0.592)& (0.272 0.546) \\
        \multirow{2}{*}{$Y_2$} & Est.        &  -2.802& -0.842 &      -0.452  \\
        & (CI) &  (-3.038 -2.541)& (-0.856 -0.826)& (-0.498 -0.405 ) \\
        \multirow{2}{*}{$Y_3$} & Est.        &  8.256&0.040&   -0.736      \\
        & (CI) & (6.891 9.607)& (-0.042  0.112)& (-0.985 -0.570) \\
        \hline          \hline
    \end{tabular}   \caption{Point estimates of the regression coefficients$\times 1000$ at each ecoregion and for each Gaussian process component. Ecoregions are coded as follows $1=1A$,
$2=1B$, $3=1C$, $4=1D$, $5=2A$, $6=2B$ and $7=2C$.} \label{tab:ecobeta2}
\end{table}

After model fitting, we used posterior samples to predict the values of the three variables over a $15Km \times 15km$  grid of 3305 spatial
points over 720 times. Posterior estimates of each time series were obtained in only 20 minutes on the TeraStat cluster \citep{dsscluster}
that allows for fast computing with a limitation on the number of processes that can be lunched simultaneously. Such limitation is not
implemented at the Bari ReCaS DataCenter that provides a computing power of 128 servers each with 64 cores and 256GB of RAM. It houses a
small cluster dedicated to High Performance Computing, running applications using many cores at the same time, that was used for parallel
computing of the predictions. To asses the out of sample predictive capability of the chosen model, we built a validation set starting with
the finest definition of ecoregions ($k=5$, 35 ecoregions). We selected 10$\%$ of the available observations with at least one station per
ecoregion, excluding ecoregions with only one station (1A1a,1D1a). The validation set was used to evaluate the root mean squared prediction
error for each response variable on its original scale, obtaining very encouraging results: Rain 5.8mm, T. min 1.3$^\circ$, T. max
1.4$^\circ$ corresponding to a relative error\footnote{$100\cdot RMSE/Range(Y)$} of 0.38\% for the Rain, 2.64\% and 2.47\% for the maximum and minimum temperature respectively. 


\section{Concluding remarks and future developments}\label{cyc:conclusion}

In this paper we present a multivariate generalization of the NNGP model proposed in \citet{datta2016a}. Our model combines the
computational efficiency of NNGP's with several new ideas for handling complex structures typical of climate variables. We use the linear
model of coregionalization to account for multivariate spatio-temporal dependencies, a circular representation of the time index to define
the annual cycles and propose an efficient implementation of the NNGP that allows to estimate the model with a huge amount of data. Results
are very encouraging as we are able to interpolate climate variables at any spatial and temporal resolution and impute missing values. The
richness of model output allows to characterise the Italian ecoregions with respect to rainfall, minimum and maximum temperature returning
information on cyclical trend, spatial and temporal correlation.

The future will find us working on more detailed bioclimatic characterisation of the Italian ecoregions. To do that we need to
obtain parameter estimates for all the available ecoregional tiers, including Divisions, Sections and Subsections. We are interested in better understanding the role of climate variables at a more detailed ecoregion level. 
  Further, as new ecoregional boundaries have recently been proposed mainly based on biogeographic and physiographic considerations (Blasi et al., unpublished data), the model could  be
  applied to develop a climatic characterisation of the new strata, comparing results to those reported in this paper.

\section*{Acknowledgments}
The work of the first three authors  is partially developed under the PRIN2015 supported-project ‘‘Environmental processes and human activities: capturing their interactions via statistical methods (EPHASTAT)’’ funded by MIUR (Italian Ministry of Education, University and Scientific Research) (20154X8K23-SH3).
\bibliographystyle{natbib}
\bibliography{all}


\end{document}